\documentclass[10pt, aps, prd, amsmath, amssymb, floats, floatfix, twocolumn, notitlepage, superscriptaddress, nofootinbib, showpacs, longbibliography]{revtex4-2}
\usepackage[T1]{fontenc}
\usepackage[utf8]{inputenc}
\usepackage{lmodern}
\usepackage{mathbbol}
\usepackage{verbatim}
\usepackage{physics}
\usepackage{orcidlink}

\definecolor{linkcolor}{rgb}{0.6, 0.0, 0.0}
\usepackage{aas_macros}
\usepackage[all]{hypcap}
\usepackage{graphicx}
\usepackage{xspace}
\usepackage{amssymb}
\usepackage{amsmath}
\usepackage[normalem]{ulem} 
\usepackage{bm} 
\usepackage{microtype}
\usepackage[english]{babel}
\usepackage{blindtext}
\usepackage{array}
\usepackage{tabularx}
\usepackage{multirow}
\usepackage{subfigure}
\usepackage{natbib}
\graphicspath{%
  {figs/}%
}

\usepackage{cleveref}
\crefname{section}{Section}{Sections}
\crefname{equation}{Eq.}{Eqs.} 
\crefname{figure}{Fig.}{Figs.}
\crefname{appendix}{Appendix}{Appendices}
\usepackage{soul}
\allowdisplaybreaks
\definecolor{db}{rgb}{0.0, 0.0, 0.62}
\definecolor{dm}{rgb}{0.7, 0.01, 0.7}
\definecolor{dr}{rgb}{0.55, 0.0, 0.0}
\newcommand{\lt}{\left}
\newcommand{\rt}{\right}
\newcommand{\p}{\partial}
\begin{document}
\title{Tidal deformation of an accreting compact object}

\author{Avijit Chowdhury\orcidlink{0000-0002-7235-5076}}
\email{avijit.chowdhury@iiap.res.in}
\affiliation{Indian Institute of Astrophysics, Block 2, 100 Feet Road, Koramangala, Bengaluru 560034, India}

\affiliation{Department of Physics, Indian Institute of Technology Guwahati, Assam-781039, India}

\author{Chiranjeeb Singha\orcidlink{0000-0003-0441-318X}}
\email{chiranjeeb.singha@iucaa.in}
\affiliation{Inter-University Centre for Astronomy and Astrophysics, Post Bag 4,   Pune 411 007, India}

\author{Kazuharu Bamba\orcidlink{0000-0001-9720-8817}}
\email{bamba@sss.fukushima-u.ac.jp}
\affiliation{Faculty of Symbiotic Systems Science, Fukushima University, Fukushima 960-1296, Japan}

\author{Sumanta Chakraborty\orcidlink{0000-0003-3343-3227}}
\email{tpsc@iacs.res.in}
\affiliation{School of Physical Sciences, Indian Association for the Cultivation of Science, Kolkata-700032, India}

\begin{abstract} 
Tidal deformation of a compact object serves as a sensitive probe of the strong-gravity regime and the nature of the compact object. It captures how a compact object responds to the external perturbing field of a companion. In realistic astrophysical settings, compact objects are typically immersed in matter-rich environments, which can significantly alter the response. In this work, we investigate the static deformability of Schwarzschild-like exotic compact objects (ECOs) embedded in a quasi-stationary, self-gravitating thin accretion disk. By modelling the external spacetime with a relativistic thin-disk solution, we isolate environmental contributions to the scalar and spin-1 response while maintaining analytical control. We show that, for perfectly reflecting ECOs, the characteristic logarithmic dependence of the scalar and spin-1 response on compactness, set by near-horizon physics, remains intact even in the presence of accretion. The disk primarily amplifies the overall magnitude of the response significantly. These findings highlight that environmental effects can seriously impact tidal signatures, while still permitting, under suitable conditions, the distinguishability of horizonless compact objects from black holes in gravitational-wave observations.
\end{abstract}
\maketitle
\section{Introduction}

Tidal deformation of a compact object provides a robust and physically well-defined probe of strong gravity and the nature of the deformed compact object. Roughly speaking, in a binary system, the tidal deformation characterizes how a compact object in the binary responds to the gravitational field of the companion and is computed by relating the induced field multipole moments of the deformed object to the applied tidal field. 
This response is quantified by Love numbers (LNs), first introduced in the Newtonian regime to describe the lunar and solar tides on Earth \cite{1909MNRAS..69..476L, 10.1098/rspa.1909.0008, 1911spge.book.....L}.
  
In the strong field region, the relativistic computation of the LNs, on the other hand, has been performed for the first time in Refs. \cite{Hinderer:2007mb,  Binnington:2009bb, Damour:2009vw} and these results are used routinely in gravitational-wave modelling and precision tests of gravity in the strong-field regime \cite{Flanagan:2007ix, Yagi:2016bkt, Yagi:2013bca, Yagi:2013awa, Cardoso:2017cfl, Abbott:2018exr, LIGOScientific:2017vwq, Abbott:2018wiz} (for recent reviews, see \cite{Chakraborty:2026qru, Rodriguez:2026iot}).

One of the most interesting features exhibited by the LNs is the well-established fact that black holes (BHs) in four-dimensional vacuum general relativity (GR) have identically vanishing static LNs under bosonic perturbations \cite{Damour:2009vw, Binnington:2009bb, Kol:2011vg, 2013arXiv1304.2228C, Gurlebeck:2015xpa, Chia:2020yla, Bhatt:2023zsy, LeTiec:2020bos, Charalambous:2021mea, Chakraborty:2025zyb, Chakraborty:2026qru}  
\footnote{This result can be derived from several complementary perspectives. For instance, the existence of a ladder-like symmetry, analogous to that of the quantum harmonic oscillator, can be invoked to explain the vanishing of LNs \cite{hui2022ladder-678, achour2022hidden-8c7, charalambous2021hidden-5e0}. Likewise, approaches based on effective field theory and scattering amplitudes consistently lead to the same conclusion, namely that BHs in four-dimensional vacuum general relativity possess identically vanishing LNs \cite{ivanov2023vanishing-9aa, creci2021tidal-42e, Charalambous:2021mea}.}. This property turns out to be not universal. Going beyond any one of these five assumptions: (a) four dimensions, (b) vacuum spacetimes, (c) GR as the theory of gravity, (d) asymptotically flat BHs and (e) static perturbations, generically leads to non-vanishing LNs. This has been shown for --- (a) higher and lower dimensions, in Refs. \cite{Kol:2011vg, Cardoso:2019vof, Hui:2020xxx, DeLuca:2024ufn, Bhatt:2024mvr, Chakravarti:2018vlt, Rodriguez:2023xjd}, (b) for higher curvature gravity theories, in Refs. \cite{Cardoso:2018ptl, Cardoso:2017cfl, DeLuca:2022tkm, Singha:2025xah, Barbosa:2025uau, Barbosa:2026qcv}, (c) for non-trivial asymptotic behaviour, in Refs. \cite{Emparan:2017qxd, Nair:2024mya, Franzin:2024cah}, (d) non-trivial boundary condition at/near the horizon, in Refs. \cite{Silvestrini:2025lbe, Chakraborty:2023zed, Nair:2022xfm, Pani:2015hfa, Cardoso:2017cfl}, (e) dynamical perturbations, in Refs. \cite{Chakraborty:2025wvs, Bhatt:2023zsy, Bhatt:2024yyz, katagiri2024relativistic-5c0, creci2021tidal-42e, saketh2023dynamical-b30, Chakraborty:2023zed, Kehagias:2024rtz, Chakraborty:2026dox, Perry:2024vwz, Ghosh:2026vig}, and (f) non-trivial matter background, in Refs. \cite{Chakraborty:2024gcr, Chakravarti:2025awj, Cardoso:2019upw, Cardoso:2021wlq, DOnofrio:2026ulh, Zhao:2026eti, Cannizzaro:2024fpz}. 

Among all of these, in the present work we will concentrate on the study of the effects of non-trivial boundary conditions near the horizon on the LNs in a non-trivial matter background. The non-trivial boundary condition near the horizon can potentially arise from ultra-compact objects, which have compactness close to the BHs, but do not have any event horizon. Since these objects typically violate energy conditions, they are also referred to as exotic compact objects (ECO). These objects arise as alternatives to the BH paradigm, and cure several issues associated with BH spacetimes, namely the existence of a singularity, the absence of deterministic evolution and compatibility with inclusion of quantum effects \cite{Cardoso:2016oxy, Cardoso:2016rao, Bueno:2017hyj, Cardoso:2019rvt, Maggio:2021ans, DelGrosso:2023trq, Pani:2010em, Biswas:2023ofz, Abedi:2016hgu, Dey:2020lhq, Mark:2017dnq, Chakravarti:2021clm, Biswas:2022wah, Biswas:2026zif}. As is evident from the previous discussion, a key characteristic of ECOs is the absence of an event horizon, even though their exterior spacetime geometry closely mimics that of BHs. The location of the ECO surface with an exterior Schwarzschild spacetime is typically expressed using the following relation:
\begin{align}
r_{0}=2M(1+\widetilde \epsilon)\,,
\end{align}
where $M$ is the mass of the Schwarzschild BH and $\widetilde \epsilon \ll 1$ is a dimensionless quantity that characterizes the difference in the compactness of a BH and an ECO. The static LNs of an ECO have a very interesting Logarithmic scaling with its compactness $\widetilde\epsilon$ \cite{Cardoso:2017cfl, Chakraborty:2023zed, Silvestrini:2025lbe}, and our interest is to study how the Logarithmic behaviour is modified due to the presence of an accretion disk. The presence of an accretion disk is typical for realistic astrophysical scenarios, as compact objects are seldom isolated. The accretion disk around a BH is generated by infalling matter fields possessing non-zero angular momentum, leading to a rotationally supported disk-like structure around the compact object. In realistic accretion disk models, angular momentum transport, turbulent stresses, and radiative processes drive the slow inward flow of matter and also lead to electromagnetic emission from the disk \cite{Shakura:1972te, Page:1974he, Novikov:1973kta, Chakrabarti:1996cc}. Moreover, we would also like to compare our results with those in \cite{Cannizzaro:2024fpz}, where the LNs of a BH with an accretion disk had been computed. It must also be stressed that, in this work the accretion timescale has been assumed to be much longer than all other dynamical timescales associated with the problem. Otherwise, this will lead to a pile-up of matter near the ECO surface \cite{Sharma:2026gvu}. 

In the present work, we have adopted an effective relativistic description of the matter distribution in the accretion disc, outside the surface of the ECO, using a quasi-stationary, self-gravitating thin disk approximation \cite{Kotlarik:2018nbd, Kotlarik:2022spo, Chen:2023akf}. This leads to an analytically tractable metric which captures the spacetime effects of the ECO as well as the gravitational effect of accretion on the surrounding geometry \cite{Kotlarik:2018nbd, Kotlarik:2022spo}. 
Within this setup, involving an accretion disk, the ringdown spectrum \cite{Chen:2023akf} and tidal deformabilities \cite{Cannizzaro:2024fpz} of BHs have been studied. It was found that the modifications to the external spacetime geometry of the BHs in the presence of the self-gravitating accretion disk affect these observables significantly.

Motivated by these findings, in this paper we investigate how the presence of an accretion disk around a horizon-less compact object modifies its tidal response. {We also ask when the disk contribution dominates over the intrinsic response of the ECO in a realistic binary, and whether an accreting ECO can then be distinguished from an accreting BH.}

This paper is organized as follows. In \cref{sec:BG}, we briefly review the effective geometry of a non-rotating compact object surrounded by an accretion disk. In \cref{sec:computation}, we compute the LNs for an accreting compact object. Our main results are presented in \cref{sec:result}. Finally, we conclude with a discussion on our findings and future prospects in \cref{sec:summary}. We have also added supplementary calculations regarding --- (a) details of the spacetime geometry generated by the accretion disk in \cref{metric}, and (b) some additional information used in the main text in \cref{app:psi-bh-1}. 

\textit{Notations and Conventions:}  
Throughout this paper, we adopt the mostly-plus signature convention, such that in $3+1$ dimensions the Minkowski metric in Cartesian coordinates is given by $\mathrm{diag}(-1,+1,+1,+1)$. Unless otherwise stated, we work in natural units with $G=1=c$ throughout the paper.

\section{Background Geometry}\label{sec:BG}
In this section, we briefly review the effective geometry of a non-rotating compact object surrounded by an accretion disk. We assume that the thickness of the accretion disk is much smaller than the size of the central compact object, so that the disk can be modelled as infinitesimally thin. Indeed for astrophysical BHs, the typical disk height is often negligible compared to the horizon radius \cite{Kuzmin,1963ApJ...138..385T,1993MNRAS.265..126B}. In what follows we will borrow the results of~\cite{Kotlarik:2018nbd, Kotlarik:2022spo, Chen:2023akf} and consider
a Schwarzschild BH of mass $M$, surrounded by a thin accretion disk with a characteristic radius $b\gg 2M$ and finite total mass $\mathcal{M}_{\rm d}$, referred to as the Schwarzschild BH-disk model. A very similar technique, as we discuss below, can be used to obtain the disk model for spherically symmetric and static ultra-compact objects.   

The compact object-disk spacetime is constructed within the class of static\footnote{Since the typical accretion timescale of a compact object \cite{1978A&A....63..221A,Olivares-Sanchez:2024dfh} is much longer than the orbital timescale in a compact binary, we treat the background spacetime as static. Moreover, modelling the disk as two identical counterrotating streams of pressure-less dust following circular geodesics~\cite{Kotlarik:2018nbd, Kotlarik:2022spo, Chen:2023akf} ensures the absence of net angular momentum and eliminates frame-dragging effects.}, axisymmetric Weyl solutions, in which the metric is described by two functions $\nu(\rho,z)$ and $\lambda(\rho,z)$ in Weyl coordinates $(\rho,z)$. The Weyl-type metric is given by,
\begin{equation}\label{gen_metric}
ds^2 = - e^{2\nu} d t^2 + \rho^2 e^{-2\nu} d\phi^2 + e^{2\lambda - 2\nu} (d\rho^2 + d z^2)~.
\end{equation}
The use of the Weyl coordinate system is particularly advantageous for this problem, since the substitution of \cref{gen_metric} in Einstein's equations suggests that the metric function $\nu(\rho,z)$ satisfies a linear Laplace equation in flat three-dimensional space $\nabla^{2}\nu=0$, while the other metric function $\lambda$ satisfies,
\begin{align}
\partial_{\rho}\lambda=\rho\lt[\left(\partial_{\rho}\nu\right)^2-\left(\partial_{z}\nu\right)^2\rt]\,, 
\quad 
\partial_{z}\lambda=2 \rho \partial_{\rho}\nu\partial_{z}\nu~.
\end{align}
This property allows one to exploit techniques analogous to Newtonian gravity and construct composite configurations via linear superposition at the level of $\nu$. In contrast, the second metric function $\lambda$ is determined through non-linear differential relations, encoding the genuinely relativistic interactions between different sources.

In what follows we will consider a static and spherically symmetric central compact object (may or may not be a BH), surrounded by a thin accretion disk. Thus the geometry outside the compact object is described by a non-linear superposition of the Schwarzschild spacetime with the spacetime geometry of the accretion disk. The Schwarzschild spacetime outside the compact object is represented in Weyl coordinates through the following metric functions $\nu_{\rm Schw}$ and $\lambda_{\rm Schw}$, such that \cite{Stephani:2003tm},
\begin{align}
e^{2\nu_{\rm Schw}}&=\frac{r_{+}+r_{-}-2M}{r_{+}+r_{-}+2M}
\\
e^{2\lambda_{\rm Schw}}&=\frac{(r_{+}+r_{-}-2M)(r_{+}+r_{-}+2M)}{4r_{+}r_{-}}\,,
\end{align}
where, $r_{\pm}^{2}=\rho^{2}+(\lt|z\rt|\pm M)^{2}$.
Note that we have the following relation between the Weyl coordinates $(\rho,z)$ and the standard spherically symmetric coordinate system $(r,\theta)$ for the Schwarzschild background,
\begin{equation}\label{eq:trans}
\rho = \sqrt{r(r-2M)}\,\sin\theta~, \qquad z = (r-M)\cos\theta~.
\end{equation}
As is evident, this transformation is valid only for $r>2M$, which is the case here. 

On the other hand, a physically viable thin-disk source can be obtained by starting from the Kuzmin-Toomre family of Newtonian disk solutions and performing a Kelvin inversion~\cite{Kuzmin, 1963ApJ...138..385T, 1992MNRAS.257..152E, 1993MNRAS.265..126B}. This procedure yields a thin disk model with finite total mass, whose surface density vanishes at the center, located at $\rho=0$ (or, equivalently at $r=2M$) and decays sufficiently rapidly at large radii, ensuring regularity of the spacetime everywhere outside the center and avoiding the presence of sharp edges or singular rings. The corresponding relativistic metric functions for the disk, henceforth denoted as $\lambda_{\rm disk}$ and $\nu_{\rm disk}$, can be obtained in closed analytic form within the Weyl formalism (see Appendix \ref{metric} for details). 
 
Owing to the linearity of the Laplace's equation, the total gravitational potential, described by the $1+2\nu$, arising from the $g_{tt}$ component of the metric, is given by a simple superposition,
\begin{equation}
\nu = \nu_{\rm Schw} + \nu_{\rm disk}~,
\end{equation}
whereas the other metric function $\lambda$ takes the form
\begin{equation}
\lambda = \lambda_{\rm Schw} + \lambda_{\rm disk} + \lambda_{\rm int}~,
\end{equation}
with $\lambda_{\rm int}$ encoding the non-linear interaction between the BH and the disk. It is convenient to define
\begin{equation}
\lambda_{\rm ext} \equiv \lambda - \lambda_{\rm Schw} = \lambda_{\rm disk} + \lambda_{\rm int}~,
\end{equation}
which captures all deviations from the pure Schwarzschild geometry. Combining all these, and transforming back to the Schwarzschild radial coordinate system\footnote{While the Weyl representation is ideally suited for constructing the solution, it is not the most convenient for physical interpretation, particularly near the horizon. It is therefore useful to transform the solution to Schwarzschild-like coordinates $(r,\theta)$, in which the background geometry of the central object takes its familiar form. Thereby providing a geometrically transparent description of the system, allowing one to interpret radial distances, angular structure, and near-horizon behaviour in a physically meaningful way.}, by inverting \cref{eq:trans}, the effective metric describing this spacetime can then be written as
\begin{align}\label{eq:SBHD-metric}
ds^2=&-f(r)e^{2 \nu_{\rm disk}} dt^2
+e^{2\lambda_{\rm ext}-2\nu_{\rm disk}} \frac{dr^2}{f(r)} \nonumber\\
&+r^2 e^{-2 \nu_{\rm disk}}\lt( e^{2\lambda_{\rm ext}} d\theta^2 +\sin^2\theta d\phi^2 \rt)~.
\end{align}
Therefore, the functions $\nu_{\rm disk}(r)$ and $\lambda_{\rm ext}(r)$ encode the gravitational field of the disk and its interaction with the central compact object, whose boundary is characterized by a timelike surface located at $r_{0}=2M(1+\widetilde{\epsilon})$, along with $f(r)=1-(2M/r)$. Therefore, the surface of the object simply replaces the BH horizon. Our interest lies in the ultra-compact objects, for which $0<\widetilde{\epsilon} \ll 1$, and $\widetilde{\epsilon}\to 0$ provides us the BH limit. As a consequence, the effective metric describing the background geometry in the exterior region ($r\ge r_0$) is identical to that of the Schwarzschild geometry with a superposed disk model, as depicted above. The distinguishing feature of the ultra-compact object enters solely through the boundary conditions imposed on perturbations at the surface $r=r_0$, which encode the reflective properties of the object. In particular, for $\widetilde{\epsilon}\ll 1$, as is the case here, it is safe to assume that the presence of the surface does not modify the equilibrium configuration of the thin disk, whose stress-energy tensor and associated gravitational potentials are taken to be the same as in the BH case. This allows us to isolate the effect of near-surface reflectivity on the tidal response without altering the background geometry.

Returning to the structure of the unknown functions $\nu_{\rm disk}(r)$ and $\lambda_{\rm ext}(r)$, we follow \cite{1963ApJ...138..385T, 2009MNRAS.396.1487V, Kotlarik:2022spo}, where the most general solution for both of these functions is characterised by two integers $(m,n)$ and corresponds to a power series in the ratio $(b/r_{\rm b})$. Here $b$ is a characteristic scale associated with the accretion disk, and the quantity $r_b$ is defined as,
\begin{equation}\label{definition_rb}
r_b=\sqrt{\rho^2+(|z|+b)^2}~,
\end{equation}
from which the disk potential can be expressed as
\begin{equation}
\nu_{\rm disk}^{(m,n)}=-W^{(m,n)}\sum_{j=0}^{m+n} Q_j^{(m,n)}\frac{b^j}{r_b^{j+1}}P_j\!\lt( \frac{| z |+b}{r_b} \rt)~,
\end{equation}
where the normalization constant $W^{(m,n)}$ reads,
\begin{align}
 W^{(m,n)}=(2m+1)   \begin{pmatrix}
  m+n+1/2 \\
  n\\
\end{pmatrix} \mathcal{M}_{\rm d}~.
\end{align} 
One can verify that the mass of the total disk is $\mathcal{M}_{\rm d}$. The angular dependence of the disk is encoded through the Legendre polynomials $P_j$, while the coefficients $Q_j^{(m,n)}$ are fixed by the chosen disk model,
 \begin{align}
Q_j^{(m,n)}=\begin{cases}
        \sum_{k=0}^n(-1)^k \begin{pmatrix} n\\k\\ \end{pmatrix}\frac{2^{j-k-m}(2m+2k-j)!}{(m+k-j)!(2m+2k+1)!!}\\\quad\quad\quad\quad\quad\quad~~~~~~~~~~~~~~~~~~~~~~~~~~\rm{if}\; j\leq m\\
        \sum_{k=j}^{m+n}(-1)^{k-m} \begin{pmatrix} n\\k-m\\ \end{pmatrix}\frac{2^{j-k}(2k-j)!}{(k-j)!(2k+1)!!}\\\quad\quad\quad\quad\quad\quad~~~~~~~~~~~~~~~~~~~~~~~~~~ \rm{if}\; j>m~.
\end{cases}
\end{align}
One can verify that the parameter $n$ controls the disk density profile close to the origin $\rho=0$, whereas the $m$  governs its asymptotic behaviour.  

Choosing the mass of the accretion disk to be much smaller than the BH mass, i.e., $(\mathcal{M}_{\rm d} /M)=\epsilon \ll 1$, and because of the complicated structure of the metric, we restrict our analysis to terms linear in $\epsilon$ only. Further, the contribution from $\lambda_{\rm disk}$ is second order in $\epsilon$, see \cref{metric}, and hence we approximate $\lambda_{\rm ext}\approx\lambda_{\rm int}$, where the most general $\lambda_{\rm int}$, for a generic choice of $(m,n)$, can be obtained from the following recurrence relations,
\begin{eqnarray}
    &\lambda_{\rm int}^{(0,0)}=-\frac{\mathcal{M}_{\rm d}}{r_b}\lt( \frac{r_-}{b+M}-\frac{r_+}{b-M} \rt)-\frac{2 \mathcal{M}_{\rm d} M}{b^2-M^2}~,\\
     &\lambda_{\rm int}^{(0,n+1)}= \lambda_{\rm int}^{(0,n)}+\frac{b}{2(n+1)}\frac{\p}{\p b} \lambda_{\rm int}^{(0,n)}~,\\
     &\frac{(2m+1)(2n+3)}{2m +2n +3} \lambda_{\rm int}^{(m+1,n)}=\lambda_{\rm int}^{(m,n)}-b \frac{\p}{\p b}\lambda_{\rm int}^{(m,n)}+\nonumber\\
    &\ \ \ \ \ \  \  \ \frac{4m (n+1)}{2m +2n +3} \lambda_{\rm int}^{(m,n+1)}~,
\end{eqnarray}
where, $r_{\pm}=\sqrt{\rho^2+\lt( \lt|z\rt|\pm M \rt)^2}$, as we have defined earlier. Therefore, keeping linear order in $\epsilon$ terms alone, the spacetime metric outside an ultra-compact object, surrounded by a thin disk, can be expressed as,
\begin{align}
    g_{tt}(r, \theta)&\approx -f(r)(1+2\nu_{\rm disk})\label{eq:gtt}~,\\
    g_{rr}(r, \theta)&\approx\frac{1}{f(r)}\lt( 1+2 \lambda_{\rm int}-2 \nu_{\rm disk}\rt)\label{eq:grr}~,\\
    g_{\theta \theta}(r, \theta)&\approx r^2 \lt( 1+2 \lambda_{\rm int}-2 \nu_{\rm disk}\rt)\label{eq:gthth}~,\\
     g_{\phi \phi}(r, \theta)&\approx r^2\sin^2\theta \lt(1-2 \nu_{\rm disk} \rt)\label{eq:gphph}~.
\end{align}
As is evident, the above spacetime can also be interpreted as a deformed Schwarzschild spacetime~\cite{Chen:2023akf}.
One can verify that the metric functions $\nu_{\rm disk}$ and $\lambda_{\rm int}$ have the largest magnitude on the equatorial plane ($\cos \theta=0$) and the magnitude decreases as $\cos \theta\to \pm 1$. In fact for all values of $\theta$, $\lt|\nu_{\rm disk}\rt|$ and $\lt|\lambda_{\rm int}\rt|$ are much smaller than unity. Thus it makes sense to perform a Taylor expansion of the metric functions $\nu_{\rm disk}$ and $\lambda_{\rm int}$ about $\theta=(\pi/2)$, which reads,
\begin{align}\label{eq:nu-lamda}
\nu_{\rm disk}= \epsilon \sum_{j=0}^{j_{t}}\mathcal{V}_j(r) \lt|\cos^j \theta\rt|\,; 
\quad 
\lambda_{\rm int}=\epsilon \sum_{j=0}^{j_{t}}\mathcal{L}_j(r)\lt| \cos^j \theta \rt|~.
\end{align}
In \cref{eq:nu-lamda} the weight functions $\mathcal{V}_j(r)$ and $ \mathcal{L}_j(r)$ depend on the choices of $m$, $n$ and $b$, as well as the quantity $j_{t}$ can be chosen depending on the level of approximation involved. The absolute value in each term in the expansion preserves the equatorial reflection symmetry, along with the possibility of having non-zero surface density at the equatorial plane. Henceforth, we will fix $\{m,n\}=\{0,1\}$ while keeping $b$ as a free parameter, determining the shape of the accretion disk and hence the effective metric.

We would like to emphasize one interesting feature of the background spacetime. As evident from \cref{eq:nu-lamda}, for $j_{t}\neq 0$, the background metric explicitly depends on $\theta$, and hence is not spherically symmetric, but rather axi-symmetric. For $j_{t}=0$, on the other hand, it seems that both $\nu_{\rm disk}$ and $\lambda_{\rm int}$ are non-zero and are functions of the radial coordinate $r$ alone, and hence one might tempt to argue that the background is spherically symmetric, but it is \emph{not}. This can be seen most explicitly from \cref{eq:gthth} and \cref{eq:gphph}, even though $\nu_{\rm disk}$ and $\lambda_{\rm int}$ are functions of the radial coordinate $r$ alone, $g_{\phi \phi}\neq \sin^{2}\theta g_{\theta \theta}$, which is the cleanest signature that irrespective of the value of $j$, the background spacetime is \emph{not} spherically symmetric. This will have implications for the perturbations of the metric, considered in the next section.
\section{Love numbers of accreting ECO}\label{sec:computation}

In this section we will compute the LNs of an ECO surrounded by an accretion disk, whose details we have reviewed in the previous section. Due to the complicated nature of the metric functions, we will compute the spin-0 (scalar) and spin-1 (vector) LNs while arguing their simple generalization to the gravitational LNs. In essence, the spin-0  and spin-1 sectors capture several essential features of the problem, in particular, the sensitivity of the LNs to the properties of the accretion disk. This provides strong motivation to focus on this sector. 

\subsection{Scalar perturbation}
The computation of the static scalar response function of an ECO with an accretion disk requires solving the source-free massless Klein–Gordon (KG) equation,
\begin{align}\label{eq:KG}
\Box \Phi=0~.
\end{align}
In a Schwarzschild background, the KG equation can be separated into a radial and an angular equation using the Legendre functions as the basis for the angular part. However, in the presence of accretion the KG operator develops non-trivial dependence on  $\lt\{ r,\theta\rt\}$, destroying its separability into radial and angular parts, see \cref{eq:SBHD-metric}. However, the scale of the problem corresponds to the ratio $(\mathcal{M}_{\rm d}/M)$, and for all physically relevant scenarios, this is a very small quantity. Thus the KG operator can be expanded in a perturbative manner and it yields,
\begin{align}\label{KG_eq}
\Box \Phi = \sum_{m_z=-\infty}^{\infty} e^{i m_z \phi}\, \mathcal{D}^2_{m_z} \widetilde{\Psi}_{m_z}\lt( r, \theta\rt)=0~, 
\end{align}
where $m_z$ is the azimuthal number. Since the disk is axi-symmetric $\phi$ can be separated out. Expanding $\mathcal{D}^2_{m_z}$ up to first order in $\epsilon=(\mathcal{M}_{\rm d}/M)$, we obtain 
\begin{align}
\mathcal{D}^2_{m_z}=\mathcal{D}^2_{(0)m_z}+ \epsilon  \mathcal{D}^2_{(1)m_z} 
\end{align}
with 
\begin{align}
\mathcal{D}^2_{(0)m_z}\equiv & \frac{m_z^2 }{r^2 (1-x^2)}-\frac{1}{r^2}\p_r\lt(r^2f(r)\p_r\rt)
\nonumber
\\
& -\frac{1}{r^2}\p_x\lt[\lt(1-x^2\rt) \p_x \rt]~,
\end{align}
and
\begin{align}
\mathcal{D}^2_{(1)m_z}&\equiv%
\sum_{j=0}^{j_{\rm t}}{2 |x|^j}\Bigg[\frac{m_z^2 \mathcal{L}_j(r)}{r^2 \lt(1-x^2\rt)}  
\nonumber
\\
&\qquad +\Big(\mathcal{V}_j(r)-\mathcal{L}_j(r)\Big) \mathcal{D}^2_{(0)m_z}\Bigg]~,
\end{align}
where $x=\cos\theta$ and the summation over $j$ runs up to some value $j_t$. The angular solution of the unperturbed equation: $\mathcal{D}^2_{(0)m_z}\widetilde\Psi(r,x)=0$, can be expressed in terms of the associated Legendre polynomial, $P_{\ell_{0}}^{m_z}(x)$, where $\lt|m_{z}\rt|\leq \ell_{0}\in \mathbb{Z}^{+}$. As perturbation due to the accretion disk is considered, as in quantum mechanics, this leads to mode mixing, i.e., perturbations excite angular harmonics with $\ell\neq \ell_{0}$. Hence the most general solution for the perturbed variable $\widetilde\Psi(r,x)$ can be separated into radial and angular parts as,
\begin{align}
\widetilde{\Psi}_{m_z}(r,x)&=\sum_{\ell_{0}=0}^{\infty}\Big[P^{m_z}_{\ell_{0}}(x)\,\Psi_{{\ell_{0}} m_z}(r)
\nonumber
\\
&\qquad +\epsilon\sum_{\ell\neq{\ell_{0}}}P^{m_z}_{\ell}(x)\Psi_{{\ell} m_z}(r)\Big]\,.
\end{align}
Plugging it back into the reduced KG equation in \cref{KG_eq}, we find,
\begin{align}\label{eq:KGdiag-non}
&\mathcal{D}^2_{m_z}\widetilde{\Psi}_{m_z}(r,\theta)
\nonumber
\\
&=\sum_{\ell_{0}=0}^{\infty}\Big\{\lt(\mathcal{D}^2_{(0)m_z}+\epsilon  \mathcal{D}^2_{(1)m_z}\rt)\lt[P^{m_z}_{\ell_{0}}(x)\,\Psi_{{\ell_{0}} m_z}(r)\rt] 
\nonumber
\\
&\qquad +\epsilon \sum_{\ell\neq{\ell_{0}}}\mathcal{D}^2_{(0)m_z}\,\lt[P^{m_z}_{\ell}(x)\Psi_{{\ell} m_z}(r)\rt]\Big\}=0\,.
\end{align}
The off-diagonal terms in the third line of \cref{eq:KGdiag-non} can be removed by taking an inner product of the complete expression with $P^{m_z}_{\ell_{0}}(x)$. This yields, 
\begin{align}\label{eq:D01-int}
\! \! &\int_{-1}^1 dx P^{m_z}_{\ell_{0}}(x)\mathcal{D}^2_{m_z}\widetilde{\Psi}_{m_z}(r,x)=\mathcal{I}^{(0)}+\epsilon \mathcal{I}^{(1)}=0\,,
\end{align}
where,
\begin{align}\label{eq:D0-int}
&\mathcal{I}^{(0)}=\int_{-1}^1 dx P^{m_z}_{\ell_{0}}(x)\mathcal{D}^2_{(0)m_z}\lt[ P^{m_z}_{\ell_{0}}(x)\Psi_{{\ell_{0}} m_z}(r) \rt]
\nonumber 
\\
&=\frac{1}{r^2}\lt[\ell_{0}(\ell_{0}+1)-\p_r\lt(r^2f(r)\p_r\rt)\rt]\Psi_{\ell_{0} m_z}\mathcal{N}_{\ell_{0}m_z}\,,
\end{align}
and at first order we have,
\begin{align}\label{eq:D1-int}
&\frac{1}{\mathcal{N}_{\ell_{0} m_z}}\int_{-1}^1 dx P^{m_z}_{\ell_{0}}(x)\mathcal{D}^2_{(1)m_z}\lt[ P^{m_z}_{\ell_{0}}(x)\Psi_{{\ell_{0}} m_z}(r) \rt]
\nonumber 
\\
&= \frac{2}{r^2}\sum_{j=0}^{j_{t}}\Big\{\mathcal{L}_j (r) a^j_{\ell_{0}m_z} +\lt[\mathcal{V}_j(r)-\mathcal{L}_j(r)\rt]b^j_{\ell_{0}m_z}
\nonumber
\\ 
&\times \lt[\ell_{0}(\ell_{0}+1)-\p_r\lt(r^2f(r)\p_r\rt)\rt]\Big\}\Psi_{\ell_{0} m_z}(r)~.
\end{align}
In arriving at the final results, we have used the orthonormality of the associated Legendre functions,
\begin{align}
\int_{-1}^{1} P^{m_z}_{\ell}(x)P^{m_z}_{\ell'}(x) dx=\underbrace{\frac{2\lt(\ell+m_z\rt)!}{(2\ell+1)\lt(\ell-m_z\rt)!}}_{\mathcal{N}_{\ell m_{z}}}\delta_{\ell,\ell'}~,
\end{align}
as well as have introduced the following definitions, 
\begin{align}
&a^j_{\ell m_z}\equiv\frac{2m_z^2}{\mathcal{N}_{\ell m_z}}\int_{0}^1 \frac{x^j \lt( P^{m_z}_\ell\rt)^2}{1-x^2} dx\,,
\\
&b^j_{\ell m_z}\equiv\frac{2}{\mathcal{N}_{\ell m_z}}\int_{0}^1 x^j \lt( P^{m_z}_\ell\rt)^2 dx\,,
\end{align}
along with the Legendre equation. For the subsequent discussion, the multipole number $\ell_0$ in \cref{eq:D0-int,eq:D1-int} will be replaced by $\ell$ for brevity, since the off-diagonal terms have already been excised. 

Taking cognizance of \cref{eq:D01-int,eq:D0-int,eq:D1-int}, we arrive at the following form of the radial KG equation in the presence of small spacetime deformation, i.e., upto linear order in the perturbation variable $(\mathcal{M}_{d}/M)$,
\begin{align}\label{eq:master}
&\sum_{j=0}^{j_{t}}\Big\{\frac{1+2 \epsilon b^j_{\ell m_{z}}\lt\{\mathcal{V}_j(r)-\mathcal{L}_j(r)\rt\}}{r^2}\Big[\ell(\ell+1)
\nonumber 
\\
&- \p_r\lt(r^2f(r)\p_r\rt)\Big]\Psi_{\ell m_{z}}(r)+\frac{2\epsilon}{r^2}\mathcal{L}_j (r)a^j_{\ell m_{z}}\Psi_{\ell m_{z}}(r)\Big\}=0~.
\end{align}
So far we have kept the radial function unperturbed. To compute the LNs, we now expand the radial function $\Psi_{\ell m_{z}}(r)$ upto linear order in the perturbation variable $\epsilon$ as,
\begin{align}\label{eq:psi-decom}
\Psi_{\ell m_{z}}(r)=\Psi^{(0)}_{\ell m_{z}}(r)+ \epsilon \Psi^{(1)}_{\ell m_{z}}(r)~.   
\end{align}
Thus, the radial perturbation variable at $\mathcal{O}(\epsilon^{0})$ reads,
\begin{align}
&f(r)\p^2_r \Psi^{(0)}_{\ell m_{z}}(r)+\lt(\frac{2f(r)}{r}+ f'(r)\rt)\p_r \Psi^{(0)}_{\ell m_{z}}(r)
\nonumber
\\
&-\frac{\ell(\ell+1)}{r^2}\Psi^{(0)}_{\ell m_{z}}(r)\equiv \mathbb{D}_{\rm s}\Psi^{(0)}_{\ell m_{z}}(r)=0\,.
\label{eq:psi0}
\end{align}
Using the differential operator $\mathbb{D}_{\rm s}$, as defined above, the radial equation at $\mathcal{O}(\epsilon)$ becomes,
\begin{align}
\mathbb{D}_{\rm s}\Psi^{(1)}_{\ell m_{z}}(r)&=\sum_{j=0}^{j_{t}}\frac{2}{r^2}\mathcal{L}_j (r) a^j_{\ell m_{z}}\Psi^{(0)}_{\ell m_{z}}(r)
\nonumber
\\
&\equiv S_{\ell m_{z}}[\Psi^{(0)}_{\ell m_{z}}]\,.
\label{eq:psi1}
\end{align}
The zeroth order equation, as in \cref{eq:psi0} is simply the radial part of the KG equation on a Schwarzschild background, while the effect of accretion comes into \cref{eq:psi1}. The first-order equation has a source term dependent on $a^j_{\ell m_{z}}$, which depends on the azimuthal number $m_{z}$. For example, in the $j_{t}=0$ case, one obtains $a^{0}_{\ell m_{z}}=m_z\{(2\ell+1)/2\}$. The dependence of the radial perturbation equation on the azimuthal number $m_{z}$ is directly related to the absence of spherical symmetry for the background spacetime. Further, in the $j_t=0$ case, the radial function $\mathcal{L}_{0}$ has the following behaviour,
\begin{align}\label{eq:L0}
&\mathcal{L}_0(r)=\lt(b^2-M^2\rt)^{-2}\lt\{b^2+r (r-2 M)\rt\}^{-3/2}
\nonumber
\\
&\times \Big[b^4 M^2 (M-r)
+b^2 M^4 \lt(2 \sqrt{b^2-2 M r+r^2}+M-r\rt)
\nonumber
\\
&+2 M^4 r (r-2M) \lt(\sqrt{b^2-2 M r+r^2}+M-r\rt)\Big]\,.
\end{align}
We note that $\mathcal{L}_0(r)$ remains negative for all $r \in [2M, \infty)$. 
Due to the dependence of the source term in \cref{eq:psi1} on the azimuthal number $m_{z}$, it follows that the equation satisfied by $\Psi^{(1)}_{\ell m_{z}}$ is identical to the zeroth-order equation, for $m_{z}=0$. Since the Schwarzschild BH has zero LNs \cite{Damour:2009vw, Binnington:2009bb, Kol:2011vg,2013arXiv1304.2228C, Gurlebeck:2015xpa, Chia:2020yla, Bhatt:2023zsy}, which follows from the zeroth-order perturbation equation, it is apparent that for $m_{z}=0$, the LNs of the Schwarzschild BH with accretion disk vanish as well. Thus non-trivial results are expected only for $m_{z}>0$. This suggests that the lowest lying mode $\ell=0=m_{z}$, for the scalar field, does not contribute to the LN. Thus we need to consider higher modes. In this work, we mainly concentrate on the $\ell=2=m_{z}$ case, while we will also discuss the $\ell=1=m_{z}$ case in a few occasions, for comparison. Unless explicitly mentioned, we will be quoting results for the $\ell=2=m_{z}$ case. 

We should emphasize that the zeroth-order equation can be rewritten in the Schr\"{o}dinger-like form by introducing the tortoise coordinate, as well as by redefining the perturbation variable $\Psi^{(0)}_{\ell m_{z}}$. The same procedure can also be carried out in the presence of accretion, as demonstrated in \cite{Chen:2022ynz}. However, in the present context, we will continue to use the standard radial coordinate $r$. 

\subsubsection{Schwarzschild BH with accretion disk}

Interestingly, for $j_t=0$, the static scalar LNs of a Schwarzschild BH can be computed analytically, see \cite{Cannizzaro:2024fpz}. In this context, the solution to the zeroth-order equation, as in \cref{eq:psi0}, must be regular at the horizon, and is given by,
\begin{align}\label{eq:psi0j0-sch}
\Psi^{(0)}_{22}(r)=\frac{\mathcal{A}\lt(6 r^2-12Mr+4M^{2}\rt)}{4M^2}\,,
\end{align}
where $\mathcal{A}$ is an arbitrary constant. Substituting this into \cref{eq:psi1}, and by solving the differential equation, as well as by imposing regularity at the horizon, we obtain the resulting expression for $\Psi^{(1)}(r)$, which is rather lengthy and is presented in \cref{app:psi-bh-1}. Expanding the complete solution (upto linear order in $\epsilon$), $\Psi(r)=\Psi^{(0)}(r)+\epsilon \Psi^{(1)}(r)$ near spatial infinity, we extract the static scalar LN from the ratio of the coefficients of $r^{-3}$ and $r^2$ as,
\begin{align}\label{LoveBHacc}
k_{22}^{\rm BH+Acc}=\frac{\epsilon}{11520(2 \mathbb{b}+1)^2}&\big(30720 \mathbb{b}^5+9120 \mathbb{b}^4-16992 \mathbb{b}^3 
\nonumber 
\\
&-6972 \mathbb{b}^2+796 \mathbb{b}+439\big)\,,
\end{align}
where, $\mathbb{b}\equiv (b/2M)$ is a dimensionless quantity. For large $\mathbb{b}$, we obtain,
\begin{align}\label{eq:LoveBHacc-large}
k_{22}^{\rm BH+Acc}=\epsilon\lt(\frac{2 \mathbb{b}^3 }{3}-\frac{15 \mathbb{b}^2}{32}-\frac{ \mathbb{b} }{15}+\frac{25}{768}\rt)\,.
\end{align}
We note that the LN is linear in the expansion parameter $\epsilon$ and it has a leading behaviour $\sim\mathbb{b}^3$. If the LNs are instead computed with the solution of the radial equation in the tortoise coordinate, one obtains the dominant contribution $\sim \mathbb{b}^4$ \cite{Cannizzaro:2024fpz}. This is due to the redefinition of the master function required to cast the resulting equation into a Schrödinger-like form in tortoise coordinates. This difference in scaling is inherent when the system has an additional length scale, whether it be from the environment or a modification in the underlying gravity theory. For $j_t>0$, the linear order equation cannot be solved analytically. By employing a modified version of the methodology introduced in \cite{Cannizzaro:2024fpz} (see \cref{sec:eco-dsk-tln}), we obtain qualitatively similar behaviour of the LNs with respect to the spacetime and disk parameters as reported in \cite{Cannizzaro:2024fpz}. This serves as a sanity check for the methodology. For the sake of brevity, we refrain from reproducing the results here.

\subsubsection{Schwarzschild-like ECO with accretion disk}\label{sec:eco-dsk-tln}

In this section, we present the static scalar LNs of an accreting ECO, which form the central focus of this work. We first review the static scalar LNs of an isolated ECO before including accretion-disk induced effects. In the case of an ECO, in the dynamical situation, there would be a non-trivial reflection of the incoming radiation by the object's surface, which is encoded by the reflectivity $\mathcal{R}$ of the ECO surface, located at $r=r_{0}=2M(1+\widetilde{\epsilon})$. A perfectly absorbing surface, i.e., $\mathcal{R}=0$ corresponds to the fact that the compact object is a BH. The existence of a non-trivial reflectivity leads to modifications in the boundary conditions for perturbations, which, in turn, affect the induced multipole moments and hence the resulting LNs. At this stage, we should mention that the reflectivity can only be defined in a dynamical situation, while here we are interpreting the reflectivity to arise from the zero frequency limit of the dynamical case. In the static case, as one can anticipate from \cite{Chakraborty:2023zed}, the reflectivity takes only two possible values $\mathcal{R}=\pm 1$. For $\mathcal{R}=1$, we have a Dirichlet boundary condition $\Psi_{\rm ECO}(r_0)=0$, whereas for $\mathcal{R}=-1$, one arrives at the Neumann boundary condition, $\p_r\Psi_{\rm ECO}(r_0)=0$. As we will see, the reflectivity of the ECO surface plays a crucial role in determining its scalar response. Consequently, the reflectivity parameter provides a useful phenomenological handle for characterizing ECOs and assessing the imprint of their microscopic surface properties on observable gravitational-wave signatures. 

Having described the possible choices for the reflectivity, we start by solving the perturbation equation at the zeroth order, given by \cref{eq:psi0}, we find,
\begin{widetext}
\begin{align}\label{eq:psi0ECO}
\Psi^{(0)}_{22}|_{\rm ECO}(r)&=-\frac{\mathcal{B}}{4M^2}\Bigg\{6r^2-12M r+4M^2+\Big[\widetilde{\epsilon} \left(\widetilde{\epsilon}+1\right)\big\{\big(6r^2-12Mr+4M^2\big)\big[\log r-\log (r-2M)\big]+6M\left(2M-2 r\right)\big\} 
\nonumber
\\
&\times \big\{12M(\mathcal{R}+1)\widetilde{\epsilon}^2+6\widetilde{\epsilon} \big[\mathcal{R}(2M+2)+2M-2\big]+2M\mathcal{R}+6 \mathcal{R}+2M-6\big\}\Big]
\Big[2\widetilde{\epsilon}(\widetilde{\epsilon}+1)\coth^{-1}(2 \tilde{\epsilon}+1) 
\nonumber
\\
&\times\big\{12M(\mathcal{R}+1)\widetilde{\epsilon}^2+6\widetilde{\epsilon}\big[\mathcal{R}(2M+2)+2M-2\big]+2M\mathcal{R}+6\mathcal{R}+2M-6\big\}
\nonumber
\\
&-\mathcal{R} \big[3\widetilde{\epsilon}(\widetilde{\epsilon}+1)\big\{4M\widetilde{\epsilon}+2M+4\big\}+1\big]-3\widetilde{\epsilon}(\widetilde{\epsilon}+1)\left(4M \widetilde{\epsilon}+2M-4\right)+1\Big]^{-1}\Bigg\}~,
\end{align}
\end{widetext}
where $\mathcal{B}$ is an arbitrary constant and for $\mathcal{R}=1$, we have imposed Dirichlet boundary condition and for $\mathcal{R}=-1$, we have used Neuman boundary condition. Expanding the complete solution in \cref{eq:psi0ECO} near spatial infinity, we compute the coefficients of the corresponding growing and decaying terms, to extract the static scalar LN for the $\ell=2=m_{z}$ mode. 
For perfectly reflecting ECOs with $\mathcal{R}=\pm 1$, i.e., imposing Dirichlet/Neumann boundary conditions, the leading order LN has a characteristic Logarithmic behaviour for $\mathcal{R}=1$, such that,
\begin{align}\label{eq:K2ECO}
k_{22}^{\rm ECO}=\frac{1}{360(3+\log\widetilde{\epsilon})}\delta_{\mathcal{R},1}+\mathcal{O}(\widetilde{\epsilon})~,
\end{align}
where $\delta_{\mathcal{R},1}=1$ for $\mathcal{R}=1$ and  $\delta_{\mathcal{R},1}=0$ otherwise. This is consistent with the previous results in the literature \cite{Cardoso:2017cfl, Chakraborty:2023zed}. This suggests that for $\mathcal{R}=-1$, the static scalar LN appears only at $\mathcal{O}(\widetilde{\epsilon})$ and has no Logarithmic behaviour.
Whereas, the static scalar LN for the $\ell=1=m_{z}$ mode of an isolated Schwarzschild-like ECO is given by,
\begin{align}\label{eq:K2ECO1}
k_{11}^{\rm ECO}=\frac{1}{24(2 +\log\widetilde{\epsilon})}\delta_{\mathcal{R},1}+\mathcal{O}(\widetilde{\epsilon})~.
\end{align}
This suggests that the static LNs of Schwarzschild-like ECOs exhibit a Logarithmic behaviour in the compactness parameter $\widetilde{\epsilon}$ \emph{only} for the $\mathcal{R}=1$ case, a result which is consistent with the findings of \cite{Chakraborty:2023zed}. While, the contribution from the Neumann boundary condition always appears at $\mathcal{O}(\widetilde{\epsilon})$ and has \emph{no} Logarithmic dependence. 

Due to the existence of non-zero LNs for Schwarzschild-like ECOs, unlike the case of an accreting Schwarzschild BH, the static LNs of an accreting ECO cannot be computed analytically, even for $j_t=0$; therefore, we adopt the computational procedure detailed below. We begin by numerically solving \cref{eq:psi1} near the ECO surface to obtain a solution satisfying either the Dirichlet or the Neumann boundary condition. 

In the far away region, $\lt(r\gg r_0\rt)$, on the other hand, we consider an ansatz for the first order perturbation associated with a generic $\ell$ mode of the following form,
\begin{align}\label{eq:ansatz}
&\Psi^{(1)}_{\ell m_{z}}|_{\rm ECO} \approx~\Psi^{(0)}_{\ell m_{z}}|_{\rm ECO}
\nonumber
\\
&+C_0 \lt(\frac{r}{2M}\rt)^\ell\Bigg[1+\sum_{i=1}^{2\ell} C_i \lt(\frac{2M}{r}\rt)^i\Bigg]
\nonumber
\\
&+D_0\lt(\frac{2M}{r}\rt)^{\ell+1}\Bigg[1+D_L \log\lt(\frac{r}{2M}\rt) + \sum_{i=1}^N D_i\lt(\frac{2M}{r}\rt)^i \Bigg]~,
\end{align}
where $N$ corresponds to the truncation order of the decaying terms \footnote{In the following analysis $N$ is taken to be $10$.}. The series in the \cref{eq:ansatz}, contains both the growing and the sub-leading fall-off behaviour at large radial distances, along with Logarithmic terms. Plugging the ansatz in \cref{eq:ansatz} back into \cref{eq:psi1}, we expand it at large $r$ and solve it order-by-order for the coefficients. This determines all the coefficients in the above ansatz in terms of the coefficients $C_{0}$ and $D_{0}$, which we then determine by matching the far-zone ansatz and its derivative to the near-zone solution obtained numerically, in the intermediate region.

Using the coefficients so computed, one obtains the static scalar LN for the $\ell=2=m_{z}$ mode of an accreting ECO, which, upto leading order in $\epsilon$ and $\widetilde\epsilon$, has the approximate form,
\begin{align}\label{eq:k2ecoR1acc}
k_{22}^{\rm ECO+Acc}\approx k_{22}^{\rm ECO} - \frac{\epsilon}{6} \Big[&\frac{D_0}{2\mathcal{B}}\lt\{1+D_L \log\lt(\frac{r}{2M}\rt)\rt\}
\nonumber 
\\
& -\frac{C_0}{\mathcal{B}}~ k_{22}^{\rm ECO}\Big]+\mathcal{O}(\epsilon^2,\epsilon\tilde\epsilon)~,
\end{align}
where, the constants $D_{0}, D_{\rm L}$ and $C_{0}$ depends on the scale of the accretion disk $\mathbb{b}$, which we will determine subsequently. To ensure the robustness of the procedure, we ascertain that the obtained values are independent of the choice of the matching radius. For $j_t=0$, both in the case of an accreting ECO and an accreting BH, $D_L$  vanishes identically. However, for $j_t>0$, for both accreting BH and accreting ECO, $D_{\rm L}\neq 0$, and hence the logarithmic term persists, resulting in a sub-dominant logarithmic dependence of the computed LNs on $r$, leading to ``Running LNs''~ \cite{Barbosa:2025uau, Chakraborty:2025wvs, saketh2023dynamical-b30, goldberger2004effective-1ce, katagiri2024relativistic-5c0}.

\subsubsection{A general analysis using Green's function}\label{sec:green}

In the previous section, we have derived the LNs of an ECO surrounded by an accretion disk in \cref{eq:k2ecoR1acc}, whose scaling with the size of the accretion disk $\mathbb{b}$ will be obtained numerically, for a fixed $\ell$, in a later section.  In this section, on the other hand, we will provide an analytical estimation of the scaling of the LNs with accretion disk size $\mathbb{b}$ using Green's function for the unperturbed operator $\mathbb{D}_{\rm s}$ presented in \cref{eq:psi0}. In particular, for generic $\ell$, \cref{eq:psi0} has the following solution, 
\begin{align}\label{eq:psi0split}
\Psi^{(0)}=A_{0}\,P_{\ell}(r/M-1)+B_{0}\,Q_{\ell}(r/M-1)\,,
\end{align}
where, $P_\ell$ and $Q_\ell$ are the Legendre polynomials, with $(A_{0},B_{0})$ are the arbitrary constants. 

The Legendre polynomial $P_\ell$ is regular at $r=2M$, while, the other polynomial $Q_\ell$ diverges at $r=2M$. Thus, regularity at the horizon removes the solution $Q_\ell$ for a BH. This can also be seen from \cref{eq:psi0j0-sch}, which simply is $\,P_2(r/M-1)$. Since, at large distances, $P_{\ell}$ is a growing function, BHs have vanishing LNs.

For an ECO, on the other hand, the Dirichlet condition at $r_0$ fixes the ratio of the arbitrary constants to be $(B_0/A_0)=-(P_{\ell}/Q_{\ell})_{r=r_{0}}$, while the Neumann condition gives $(B_0/A_0)=-(\partial_{r}P_{\ell}/\partial_{r}Q_{\ell})_{r=r_{0}}$. Thus, for ECOs, there is both growing and decaying modes, as $P_{\ell}$ is a growing function, and $Q_\ell$ decays as $r^{-\ell-1}$ at large $r$. Since, $Q_{\ell}$ contains $\log(r-2M)$, it follows that for $r_{0}=2M(1+\widetilde{\epsilon})$, with $\widetilde{\epsilon}\ll 1$, we have $(B_0/A_0)\sim1/\log\widetilde{\epsilon}$, where we have considered the Dirichlet boundary condition, leading to the non-zero LNs of an isolated ECO, see \cref{eq:K2ECO,eq:K2ECO1}. Also a comparison of \cref{eq:psi0ECO} with \cref{eq:psi0split}, reveals that $A_{0}=-\mathcal{B}$. 

The first order solution $\Psi^{(1)}_{\ell m_{z}}$ has a homogeneous part, which is identical to \cref{eq:psi0split}, while the particular solution can be obtained from \cref{eq:psi1} by using the Green's function $G(r,r')$, such that, 
\begin{align}\label{eq:psi1G01}
\Psi^{(1)}_{\ell m_{z}}(r)&=A_{1}\,P_{\ell}(r/M-1)+B_{1}\,Q_{\ell}(r/M-1)
\nonumber
\\
&+\int_{r_{0}}^{\infty}dr'G(r,r')\mathbb{V}(r')\Psi^{(0)}_{\ell m_{z}}(r')\,,
\end{align}
where, $(A_{1},B_{1})$ are arbitrary constants appearing in the homogeneous solution, and the source term involves the quantity $\mathbb{V}(r)\equiv2\sum_j a^j_{\ell m_z}\mathcal{L}_j(r)$. Focussing our interest in the region $r\in [r_0,\infty)$, the Green's function $G(r,r')$ can be obtained, and hence the first order solution can be expressed as (see \cref{App:Green}),
\begin{align}\label{eq:psi1sol}
\Psi^{(1)}_{\ell m_{z}}(r)&=\left(A_{1}+\lambda_{0}A_{0}+\frac{\mathbb{I}_{\rm Q}}{M}\right)\,P_{\ell}(r/M-1)
\nonumber
\\
&+\left(B_{1}+\lambda_{0}B_{0}-\frac{\mathbb{I}_{\rm P}}{M}\right)\,Q_{\ell}(r/M-1)\,,
\end{align}
where, we have introduced the following two integrals,
\begin{align}
\mathbb{I}_{\rm Q}&\equiv\int_{r_{0}}^{r}dr'Q_{\ell}(r'/M-1)\mathbb{V}(r')\Psi^{(0)}_{\ell m_{z}}(r')
\nonumber
\\
&=A_{0}\int_{r_{0}}^{r}dr'Q_{\ell}(r'/M-1)\mathbb{V}(r')P_{\ell}(r'/M-1)
\nonumber
\\
&+B_{0}\int_{r_{0}}^{r}dr'Q_{\ell}(r'/M-1)\mathbb{V}(r')Q_{\ell}(r'/M-1)\,,
\label{integralQ}
\\
\mathbb{I}_{\rm P}&\equiv\int_{r_{0}}^{r}dr'P_{\ell}(r'/M-1)\mathbb{V}(r')\Psi^{(0)}_{\ell m_{z}}(r')
\nonumber
\\
&=A_{0}\int_{r_{0}}^{r}dr'P_{\ell}(r'/M-1)\mathbb{V}(r')P_{\ell}(r'/M-1)
\nonumber
\\
&+B_{0}\int_{r_{0}}^{r}dr'P_{\ell}(r'/M-1)\mathbb{V}(r')Q_{\ell}(r'/M-1)\,.
\label{integralP}
\end{align}
The linear-in-$\epsilon$ solution, as presented in \cref{eq:psi1sol}, must satisfy the Dirichlet condition: $\Psi^{(1)}_{\ell m_{z}}(r_{0})=0$, which yields, $(B_{1}/A_{1})=(B_{0}/A_{0})=-\{P_{\ell}(r_{0}/M-1)/Q_{\ell}(r_{0}/M-1)\}$. Thus, we have obtained the most general solution: $\Psi_{\ell m_{z}}=\Psi^{(0)}_{\ell m_{z}}+\epsilon \Psi^{(1)}_{\ell m_{z}}$, up to first order in $\epsilon$, for the scalar perturbation of an ECO immersed in an accretion disk, which reads,
\begin{align}\label{eq:scalarfull}
&\Psi_{\ell m_{z}}(r)=\left\{A_{0}(1+\epsilon \lambda_{0})+\epsilon A_{1}+\epsilon\frac{\mathbb{I}_{\rm Q}}{M}\right\}\,P_{\ell}(r/M-1)
\nonumber
\\
&+\left\{B_{0}(1+\epsilon \lambda_{0})+\epsilon B_{1}-\epsilon\frac{\mathbb{I}_{\rm P}}{M}\right\}\,Q_{\ell}(r/M-1)+\mathcal{O}(\epsilon^{2})\,.
\end{align}
The computation of the LNs requires asymptotic behaviour. Hence, we express below the large-$r$ behaviour of the Legendre polynomials. The Legendre polynomial $P_{\ell}(r/M-1)$ behaves as, $P_\ell\to p_\ell\,(r/2M)^{\ell}$, with $p_{\ell}\equiv \{(2\ell)!/(\ell!)^{2}\}$, and the other polynomial scales as, $Q_\ell\to q_\ell\,(2M/r)^{\ell+1}$, where $q_{\ell}\equiv \{(\ell!)^{2}/2(2\ell+1)!\}$. Therefore, the scalar perturbation, as presented in \cref{eq:scalarfull}, has the following asymptotic behaviour, 
\begin{align}\label{eq:scalarfullasymp}
&\Psi_{\ell m_{z}}(r)|_{\rm asymp}=\left\{A_{0}(1+\epsilon \lambda_{0})+\epsilon A_{1}+\epsilon\frac{\mathbb{I}^{\infty}_{\rm Q}}{M}\right\}\,p_{\ell}\left(\frac{r}{2M}\right)^{\ell}
\nonumber
\\
&+\left\{B_{0}(1+\epsilon \lambda_{0})+\epsilon B_{1}-\epsilon\frac{\mathbb{I}_{\rm P}^{\infty}}{M}\right\}q_{\ell}\left(\frac{2M}{r}\right)^{\ell+1}+\mathcal{O}(\epsilon^{2})\,,
\end{align}
where, $\mathbb{I}_{\rm P}^{\infty}$ and $\mathbb{I}_{\rm Q}^{\infty}$ denotes the asymptotic values of the integrals in \cref{integralP} and \cref{integralQ}. From which, we can determine the LNs by taking the ratio of the decaying to the growing coefficient, which to first order in $\epsilon$ yields,
\begin{align}\label{splitN}
&k^{\rm ECO+Acc}_{\ell m_z}=\frac{q_{\ell}}{2p_{\ell}}\frac{B_{0}(1+\epsilon \lambda_{0})+\epsilon B_{1}-\epsilon\frac{\mathbb{I}_{\rm P}^{\infty}}{M}}{A_{0}(1+\epsilon \lambda_{0})+\epsilon A_{1}+\epsilon\frac{\mathbb{I}_{\rm Q}^{\infty}}{M}}
\nonumber
\\
&=\frac{q_{\ell}B_{0}}{2p_{\ell}A_{0}}\Big[1+\epsilon \left(\frac{B_{1}}{B_{0}}-\frac{A_{1}}{A_{0}}\right)
-\epsilon \left(\frac{\mathbb{I}_{\rm P}^{\infty}}{MB_{0}}+\frac{\mathbb{I}_{\rm Q}^{\infty}}{MA_{0}}\right)\Big]
\nonumber
\\
&=\frac{q_{\ell}B_{0}}{2p_{\ell}A_{0}}\Big[1-\epsilon \frac{1}{MB_{0}}\left(\mathbb{I}_{\rm P}^{\infty}+\frac{B_{0}}{A_{0}}\mathbb{I}_{\rm Q}^{\infty}\right)\Big]\,.
\end{align}
Note that the LNs do not depend on $\lambda_{0}$, and more importantly, neither does it depend on the choices of $(A_{1},B_{1})$. Thus the LNs are solely governed by the coefficients $(A_{0},B_{0})$, as well as the integrals $\mathbb{I}_{\rm P}^{\infty}$ and $\mathbb{I}_{\rm Q}^{\infty}$. The above expression for the LNs, can be decomposed as, 
\begin{align}\label{split}
&k^{\rm ECO+Acc}_{\ell m_z}=\left(1+\alpha^{\rm extra}_{\ell m_{z}}\right)k^{\rm ECO}_{\ell m_{z}}+k^{\rm BH+Acc}_{\ell m_z}+\mathcal{O}\big(\epsilon\widetilde{\epsilon},\epsilon^2\big)\,,
\end{align}
where, the terms $k^{\rm ECO}_{\ell m_{z}}$, $k^{\rm BH+Acc}_{\ell m_z}$ and $\alpha_{\ell m_{z}}^{\rm extra}$ in \cref{split} have the following expressions\footnote{A few comments regarding the upper limit in the integrals presented in \cref{LoveBHGen} and \cref{Loveextra} are in order. The computation presented here, in the context of linear perturbation theory of an accreting compact object in the presence of an external tidal field is strictly valid only in the intermediate region $2M\ll r\ll R$, where $R$ is the distance between the deformed object and the source of the tidal field. Therefore, even if the upper limits of the integrals in \cref{LoveBHGen} and \cref{Loveextra} are spatial infinity, they should be replaced by a characteristic scale of the intermediate region, which we will choose to be $b$, the scale of the accretion disk.},
\begin{align}
&k^{\rm ECO}_{\ell m_{z}}=\frac{q_{\ell}B_{0}}{2p_{\ell}A_{0}}\,,
\label{LoveECOgen}
\\
&k^{\rm BH+Acc}_{\ell m_z}=-\frac{\epsilon q_{\ell}}{2p_{\ell}\,M}
\nonumber
\\
&\times \int_{r_{0}}^{\infty}dr'P_{\ell}(r'/M-1)\mathbb{V}(r')P_{\ell}(r'/M-1)\,,
\label{LoveBHGen}
\\
&\alpha_{\ell m_{z}}^{\rm extra}=-\frac{2\epsilon}{M}\int_{r_{0}}^{\infty}dr'Q_{\ell}(r'/M-1)\mathbb{V}(r')P_{\ell}(r'/M-1)\,
\nonumber
\\
&-\frac{2\epsilon p_{\ell}}{Mq_{\ell}}k_{\ell m_{z}}^{\rm ECO}\int_{r_{0}}^{\infty}dr'Q_{\ell}(r'/M-1)\mathbb{V}(r')Q_{\ell}(r'/M-1)\,.
\label{Loveextra}
\end{align}
One can verify that for $\ell=2=m_{z}$, the expression for $k^{\rm BH+Acc}_{22}$ coincides with the one obtained in \cref{LoveBHacc}. Thus, the expression for the LNs of a BH with an accretion disk, for generic $\ell$ is given by \cref{LoveBHGen}, up to linear order in the smallness parameter $\epsilon$. Similarly, for $\ell=2$ and for $\ell=1$, the expression for $k^{\rm ECO}_{\ell m_{z}}$ from \cref{LoveECOgen} coincides with \cref{eq:K2ECO} and \cref{eq:K2ECO1}, respectively. 

Finally, a comparison of \cref{split} with \cref{eq:k2ecoR1acc}, for the $\ell=2=m_{z}$ case, for $j_{t}=0$, fixes the following combination between the constants $D_{0}$ and $C_{0}$ to be, 
\begin{align}\label{eq:c0d0}
&\frac{MD_{0}}{24A_{0}}-\frac{MC_{0}}{12A_{0}}k_{22}^{\rm ECO}=\frac{M}{2\epsilon}k_{22}^{\rm BH+Acc}
\nonumber
\\
&-k_{22}^{\rm ECO}\int_{r_{0}}^{\infty}dr'Q_{2}(r'/M-1)\mathbb{V}(r')P_{2}(r'/M-1)
\nonumber
\\
&-\frac{p_{2}{\left(k_{22}^{\rm ECO}\right)^2}}{q_{2}}\int_{r_{0}}^{\infty}dr'Q_{2}(r'/M-1)\mathbb{V}(r')Q_{2}(r'/M-1)\,.
\end{align}
For the accreting BH, on the other hand, $B_0=0$, which implies $k_{22}^{\rm ECO}=0$, and hence the above combination fixes the constant $D_{0}$ to be,
\begin{align}\label{eq:kBHD0}
k^{\rm BH+Acc}_{22}=\frac{\epsilon\,D_0^{\rm BH}}{12A_{0}}\,.
\end{align}
Note that the above correspondence can be generalized for arbitrary $\ell$, given the structure of the LNs presented above. The above result can also be generalized for arbitrary spin, although the source term involved in the above expressions will be modified. We will discuss the implications of the above results for generic $\ell$, in the subsequent sections. 

\subsection{Spin-1 Love numbers of an accreting ECO}\label{spin-1}

In this section we compute the LNs of a spherically symmetric ECO surrounded by a thin accretion disk under an external spin-1 (vector) field perturbation. Our goal, as in the scalar case, is to determine the dependence of the corresponding LNs on the disk parameters and the ECO compactness. For simplicity, we restrict ourselves to the $j_t=0$ sector, such that the metric elements are independent of $\cos \theta$, see \cref{sec:BG}. 

\subsubsection{Asymptotic expansion and definition of Love numbers}

The LNs in the case of spin-1 perturbations are defined through the asymptotic expansion of the components $A_{t}$ and $A_{\phi}$ of the spin-1 field $A_{\mu}$, both of which are given by \cite{Chakraborty:2026qru, Cardoso:2017cfl, thorne1980multipole-8ba},
\begin{align}
A_t &= -\frac{Q}{r}+ \sum_{\ell \ge 1} \bigg[\frac{2}{r^{\ell+1}} \lt\{\sqrt{\frac{4\pi}{2\ell+1}}\, Q_\ell Y^{\ell0} + (\ell'<\ell) \rt\} 
\nonumber 
\\
&\quad -\frac{2}{\ell} r^\ell \lt\{ E_\ell Y^{\ell0} + (\ell'<\ell) \rt\}\bigg]~, 
\\
A_\phi &= \sum_{\ell \ge 1} \bigg[\frac{2}{r^\ell} \lt\{\sqrt{\frac{4\pi}{2\ell+1}}\, \frac{J_\ell}{\ell} S^{\ell0}_\phi + (\ell'<\ell) \rt\}
\nonumber 
\\
&\quad+ \frac{2}{3\ell} r^{\ell+1} \lt\{ B_\ell S^{\ell0}_\phi + (\ell'<\ell) \rt\}\bigg]~,
\end{align}
where $Q_\ell$ and $J_\ell$ are the electric and magnetic multipole moments, while $E_\ell$ and $B_\ell$ are the corresponding tidal fields with $S^{\ell m_{z}}_\phi =\sin\theta \,\partial_\theta Y^{\ell m_{z}}$.

Given the above asymptotic expansions for the `t' and `$\phi$' components of the spin-1 field, the polar (even-parity) and axial (odd-parity) LNs are defined as \cite{Cardoso:2017cfl, Chakraborty:2026qru},
\begin{align}\label{eq:S1lovedef}
k^{(s=1)}_{\ell m_{z}\rm{(E)}} &\equiv -\frac{\ell}{2R^{2\ell+1}}\sqrt{\frac{4\pi}{2\ell+1}}\, \frac{Q_\ell}{E_\ell}~, 
\\
k^{(s=1)}_{\ell m_{z}\rm{(B)}} &\equiv -\frac{3 \ell}{2 (\ell+1)R^{2\ell+1}}\sqrt{\frac{4\pi}{2\ell+1}}\, \frac{J_\ell}{B_\ell}~.
\end{align}
Here, the subscript `E/B' denotes the connection of the LNs to the polar/axial perturbations. For dimensional reasons, both the polar and axial LNs are scaled by $R^{2\ell+1}$, where $R$ is any characteristic length scale. In the present case, the characteristic scale can be chosen to be $2M$.

\subsubsection{Decomposition into polar and axial sectors}

Having derived the polar/axial LNs from the asymptotic expansions of the `t' and `$\phi$' components of the spin-1 perturbation, it is now time to decompose the generic spin-1 perturbation into axial and polar sectors. For this purpose, we first note that the dynamics of the spin-1 perturbation field $A_\mu$ is governed by the Maxwell's equations in curved spacetime,
\begin{equation}
\nabla_\nu F^{\mu\nu} = 0~,
\end{equation}
where $F_{\mu\nu} = \partial_\mu A_\nu - \partial_\nu A_\mu$. In the case of static perturbations, on the other hand, the above simply provides the master equations for axial and polar perturbations.  

In particular, we decompose the spin-1 perturbation field into vector spherical harmonic basis as~\cite{Hui:2020xxx},
\begin{align}
A_\mu &= \sum_{\ell,m_{z}} \underbrace{\left( a_0, a_r, a^{\rm (P)} \partial_\theta, a^{\rm (P)} \partial_\phi \right)}_{\rm polar} Y_{\ell m_{z}}
\nonumber
\\
&+\sum_{\ell,m_{z}} \underbrace{\left( 0, 0, \frac{a^{\rm (A)}}{\sin\theta}\partial_\phi, -a^{\rm (A)} \sin\theta \partial_\theta \right)}_{\rm axial} Y_{\ell m_{z}}~.
\end{align}
The quantities $a_0$, $a_r$, and $a^{\rm (P)}$ in the first line are functions of the radial coordinate alone and describe the polar sector. The radial function $a^{\rm (A)}$, on the other hand, describes the axial (parity-odd) sector. In what follows, we focus primarily on the quadrupolar mode $\ell=2$ and $m_z=0$. By substituting the above decomposition of the spin-1 perturbation field in terms of spherical harmonics in the Maxwell's equations, we obtain the master equations for the polar and the axial sectors, respectively. 

\subsubsection{Polar sector}

Before obtaining the master equation for the polar sector, we note that the perturbation variable $A_{\mu}$ can be transformed as $A_{\mu}\to A_{\mu}+\partial_{\mu}\Lambda$, where $\Lambda$ is a scalar, and this transformation keeps the Maxwell's equations invariant. Thus using this gauge freedom, we choose the scalar $\Lambda$, such that $a^{\rm (P)}=0$, and introduce the master variable
\begin{equation}
\Psi_{\rm (P)} = \frac{r^2}{\sqrt{6}} (a'_0 - \dot{a}_r)~,
\end{equation}
which allows us to eliminate $a_0$ and $a_r$, thereby obtaining a single differential equation for $\Psi_{\rm (P)}$ in the polar sector. In the static limit things become even simpler, as we have $a_{r}=0$ and hence the `t' component of the perturbing vector field can be expressed in terms of the master variable and its derivative, as,
\begin{equation}
a_{0}= \frac{(r-2M)\Psi_{\rm (P)}'}{\sqrt{6}r}+ \epsilon h(r)\,\Psi_{\rm (P)}~,
\end{equation}
where, the radial function $h(r)$ has the following structure,
\begin{equation}
h(r)=\sqrt{\frac{2}{3}} \frac{M(r-M)(r-2M)\lt[2M^{2}\mathbb{b}^{2}-r\lt(r-2M\rt)\rt]}{r\lt[4M^{2}\mathbb{b}^{2}+r\lt(r-2M\rt)\rt]^{5/2}}\,.
\end{equation}
Alike the case of scalar perturbation, for the polar sector of the spin-1 perturbation as well, the master equation for $\Psi_{\rm (P)}$ can be written perturbatively in the smallness parameter $\epsilon$ as,
\begin{align}
r(r-2M)\Psi^{(0)''}_{\rm (P)}+2M\Psi^{(0)'}_{\rm (P)}-6\Psi^{(0)}_{\rm (P)}&= 0~, 
\\
r(r-2M)\Psi^{(1)''}_{\rm (P)}+2M\Psi^{(1)'}_{\rm (P)}-6\Psi^{(1)}_{\rm (P)}&=S[\Psi^{(0)}_{\rm (P)}]~,
\label{spin1firstorder}
\end{align}
where the source term in the first-order-in-$\epsilon$ equation for $\Psi^{(1)}_{\rm (P)}$ is determined by the zeroth-order solution, which reads,
\begin{align}
S[\Psi^{(0)}_{\rm (P)}]=\sqrt{6}r^{2}\lt[h'(r)\Psi^{(0)}_{\rm (P)}(r)+h(r)\Psi^{(0)'}_{\rm (P)}(r)\rt]~.
\end{align}
At the zeroth order, i.e., in the absence of an accretion disk, by imposing the Dirichlet boundary condition $\lt(\Psi^{0}_{\rm (P)}(r_{0})=0\rt)$ at the ECO surface, located at $r_{0}=2M(1+\widetilde{\epsilon})$, we obtain,
\begin{widetext}
\begin{align}\label{eq:polarspin1-0}
\Psi^{(0)}_{\rm (P)}=\frac{c_{1}}{8M^{3}}\Bigg[\frac{(4 \widetilde{\epsilon}+1) (\widetilde{\epsilon}+1)^{2} \left\{2M\left(-24 r^2+12Mr+4M^{2}\right)+12r^{2}(2r-3M) \log \left(\frac{r}{r-2M}\right)\right\}}{4\left\{6\widetilde{\epsilon}(4\widetilde{\epsilon}+7)-6(4 \widetilde{\epsilon}+1)(\widetilde{\epsilon}+1)^{2}\log \left(\frac{1}{\widetilde{\epsilon}}+1\right)+17\right\}}+\left(r^3-\frac{3Mr^{2}}{2}\right)\Bigg]~,
\end{align}
\end{widetext}
where $c_{1}$ is an arbitrary constant. Note that imposing the Dirichlet boundary condition at the ECO surface is equivalent with imposing $\mathcal{R}=1$ for the ECO. In particular, from the ratio of the growing and decaying branch of the zeroth order solution in \cref{eq:polarspin1-0}, one can compute the spin-1 LNs of a spherically symmetric ECO without any accretion disk around it, which reads,
\begin{align}\label{eq:TLN-spin1-eco}
k_{2 m_z,\mathcal{R}=1}^{\rm ECO}|_{\rm polar}=\frac{3}{40(17+6\log\widetilde{\epsilon})}~.
\end{align}
We note that the above result for the static LN holds for all integer values of $m_{z}\in[-2,2]$. This is expected, as in the absence of accretion, the spacetime is spherically symmetric, leading to the tidal response being dependent on angular number alone.

In the presence of the accretion disk, on the other hand, one needs to solve for \cref{spin1firstorder} with a source term arising from \cref{eq:polarspin1-0}. For this purpose, one can use the techniques developed in \cref{sec:green}. Due to the complicated nature of the source term in \cref{eq:polarspin1-0}, it follows that a closed form analytical expression for the LNs cannot be computed in the presence of an accretion disk, and hence we evaluate them numerically, in the subsequent section, following the same procedure as described in \cref{sec:eco-dsk-tln}. 

\subsubsection{Axial sector}

The axial sector, on the other hand, is described by the variable $a^{(A)}(r)$. Again, expanding the perturbation variable, as well as the metric variables, upto linear order in $\epsilon$, we obtain the following set of perturbative equations for the axial variable,
\begin{align}
r(r-2M)a^{\rm(A,0)''}+2Ma^{\rm (A,0)'}-6a^{\rm (A,0)}&=0~, 
\\
r(r-2M)a^{\rm (A,1)''}+2Ma^{\rm (A,1)'}-6a^{\rm (A,1)} &= \, S[a^{\rm (A,0)}]~.
\end{align}
The source term, associated with the linear-in-$\epsilon$ term takes the following form,
\begin{align}
&S[a^{\rm (A,0)}]
\nonumber
\\
&=\frac{r(r-2M)(r-M)M\lt(2r^{2}-4Mr-4\mathbb{b}^{2}M^{2}\rt)}{\lt(r^{2}-2Mr +4M^{2}\mathbb{b}^{2}\rt)^{5/2}} a^{\rm (A,0)'}~.
\end{align}
As is evident, the zeroth order equation is identical for both axial and polar sectors. Thus we obtain the same axial LNs as in \cref{eq:TLN-spin1-eco} for the Schwarzschild-like ECO without an accretion disk under spin-1 perturbation. Alike the polar sector, the axial LNs of a Schwarzschild-like ECO  with an accretion disk need to be numerically evaluated, following the same procedure as described in \cref{sec:eco-dsk-tln}. We present this computation in the next section.

\section{Results}\label{sec:result}

In this section, we discuss the numerical results for LNs of both scalar and spin-1 perturbations acting on an ECO embedded in an accretion disk. The numerical results follow from the analysis described in \cref{sec:computation} and will be matched with the Green's function method, adopted in \cref{sec:green}.

\subsection{Scalar sector}

\begin{figure}
\centering
{\includegraphics[width=0.48\textwidth]{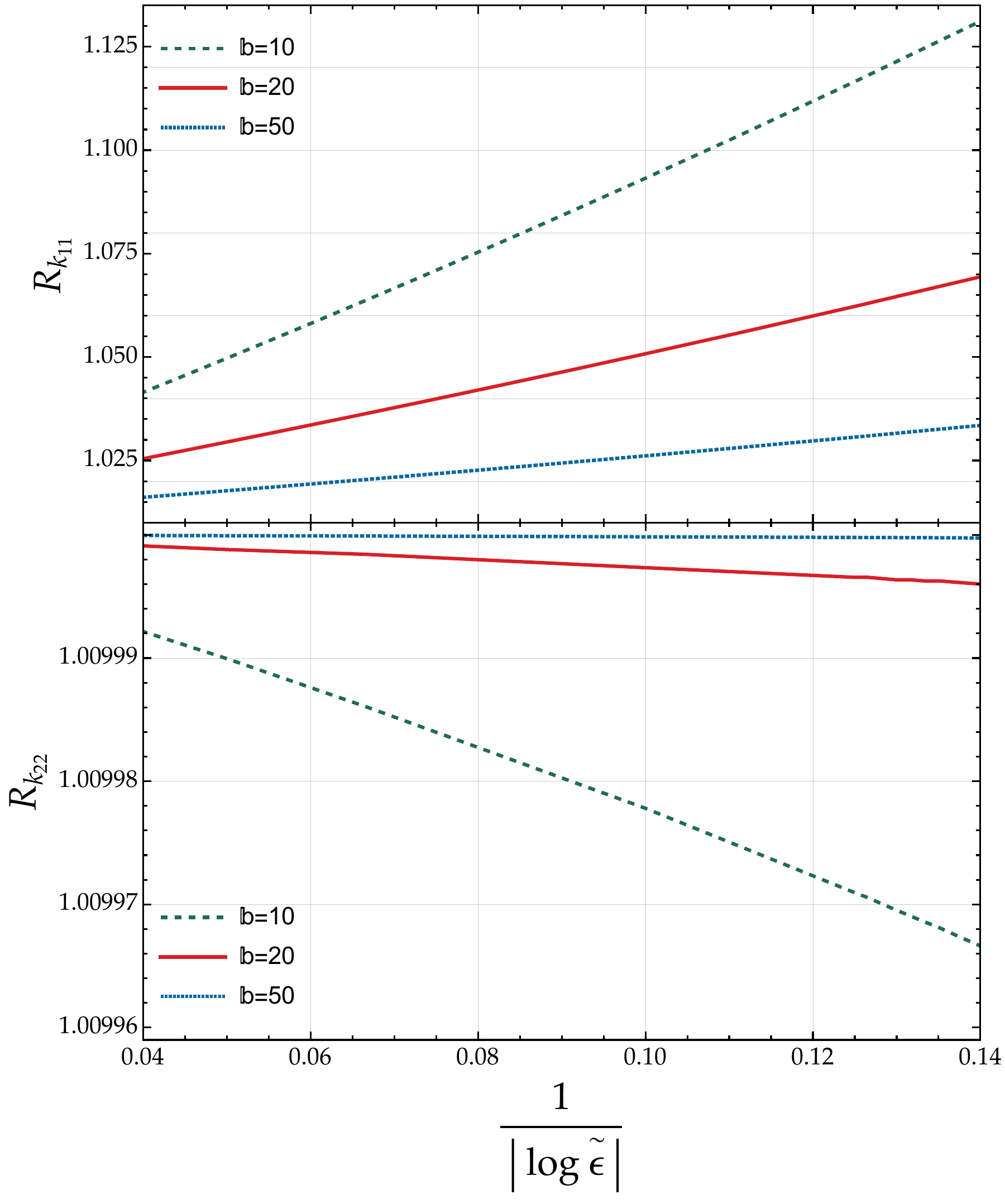}}
\caption{The ratio between the static LNs of an accreting ECO, and an accreting BH ($R_{k_{\ell mz}}\equiv k_{\ell m_{z}}^{\rm ECO+Acc}/k_{\ell m_z}^{\rm BH+Acc}$), has been plotted against $\lt|\log\widetilde{\epsilon}\rt|^{-1}$ for $\ell=m_z=1$ in the top panel and for $\ell=m_z=2$ in the bottom panel for unit reflectivity ($\mathcal{R}=1$), $\epsilon=0.01$ and fixed $\mathbb{b}$.}
\label{fig:LOmega1}
\end{figure}

To study the variation of the static scalar LNs of a perfectly reflecting accreting ECO with its compactness\footnote{Since the ECO mass is kept constant, variation of compactness of an ECO suggests a variation with respect to its surface radius $r_{0}$, which is equivalent to the variation of $\widetilde{\epsilon}$.}, we plot the ratio $(k_{\ell m_z}^{\rm ECO+Acc}/k_{\ell m_z}^{\rm BH+Acc})\equiv R_{k_{\ell m_{z}}}$ with $\lt|\log{\widetilde{\epsilon}}\rt|^{-1}$ (for $j_t=0$) in \cref{fig:LOmega1}.
As is clearly evident from \cref{fig:LOmega1}, for sufficiently compact ECOs (small values of $\widetilde{\epsilon}$) with $\mathcal{R}=1$, the ratio of static LNs between ECO with accretion and BH with accretion changes linearly with $\lt|\log\widetilde{\epsilon}\rt|^{-1}$. Thus, the characteristic Logarithmic behaviour of the static LNs for the ECOs with their compactness, see \cref{eq:K2ECO}, persists even in the presence of an accretion disk. This persistence indicates that the near-horizon structure of the ECO governs the logarithmic scaling and is not qualitatively altered by the presence of accreting matter, though the coefficient of $\lt|\log\widetilde{\epsilon}\rt|^{-1}$ depends on the scale $\mathbb{b}$, see \cref{fig:LOmega1}. This should also be evident from \cref{split}, which explicitly depicts that the ratio $R_{k_{\ell m_{z}}}$, to leading order, scales linearly with $k_{\ell m_{z}}^{\rm ECO}$, and hence must scale linearly with $\lt|\log\widetilde{\epsilon}\rt|^{-1}$, as depicted in \cref{fig:LOmega1}.

\begin{figure}[!tbp]
\centering
{\includegraphics[width=\linewidth]{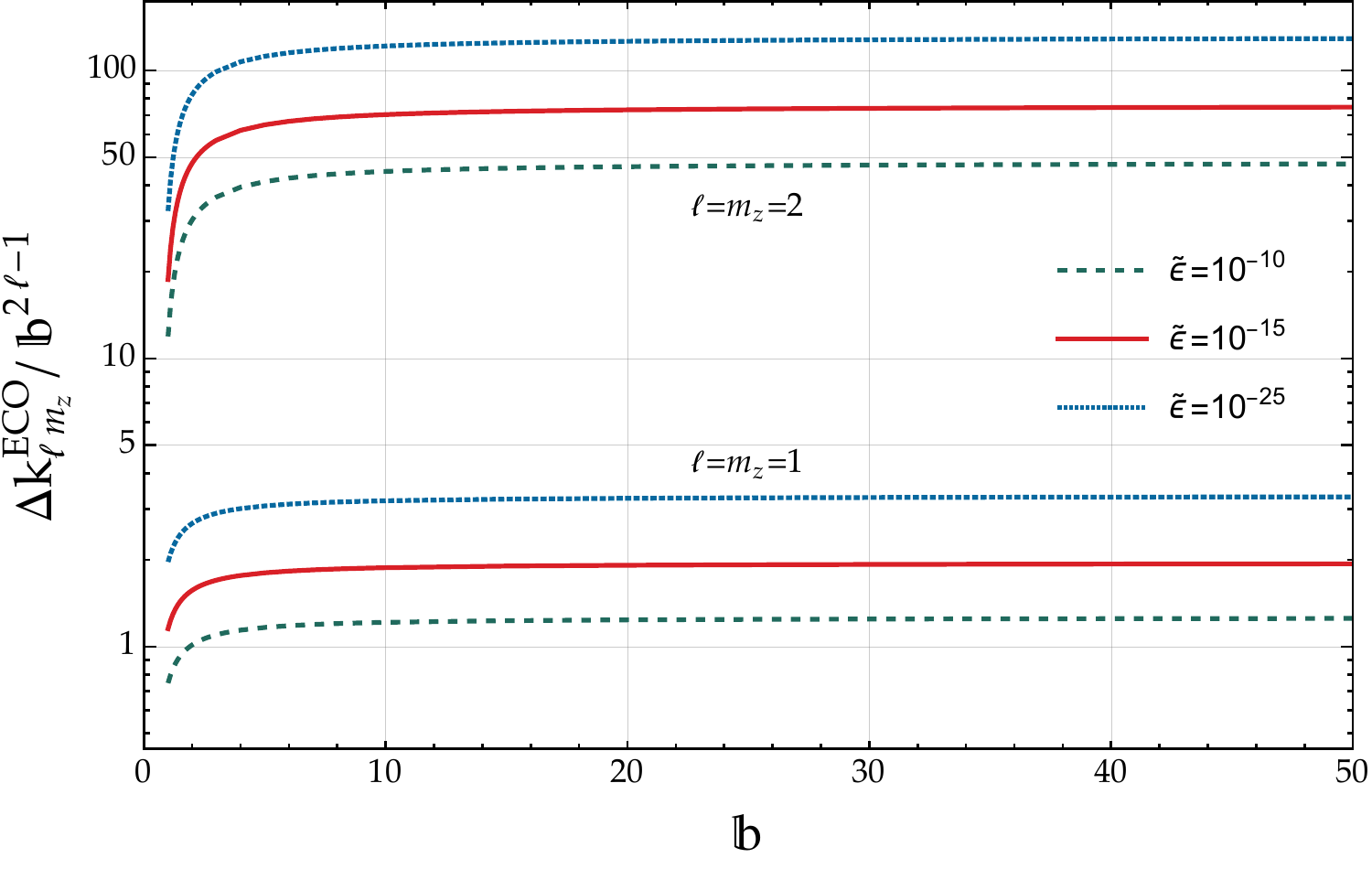}
\caption{The figure displays the quantity $(\Delta k_{\ell m_z}^{\rm ECO}/\mathbb{b}^{2\ell-1})\equiv\lt|\lt(k_{\ell m_z}^{\rm ECO+Acc}/k_{\ell m_z}^{\rm ECO}\rt)-1\rt|\mathbb{b}^{-2\ell+1}$, shown as a function of $\mathbb{b}$ for $\ell=m_z=1$ and $\ell=m_z=2$ at fixed values of $\widetilde{\epsilon}$, with $\epsilon=0.01$ and $\mathcal{R}=1$.}
\label{fig:LOmega3}}
\end{figure}

\begin{figure}[!tbp]
\centering
{\includegraphics[width=\linewidth]{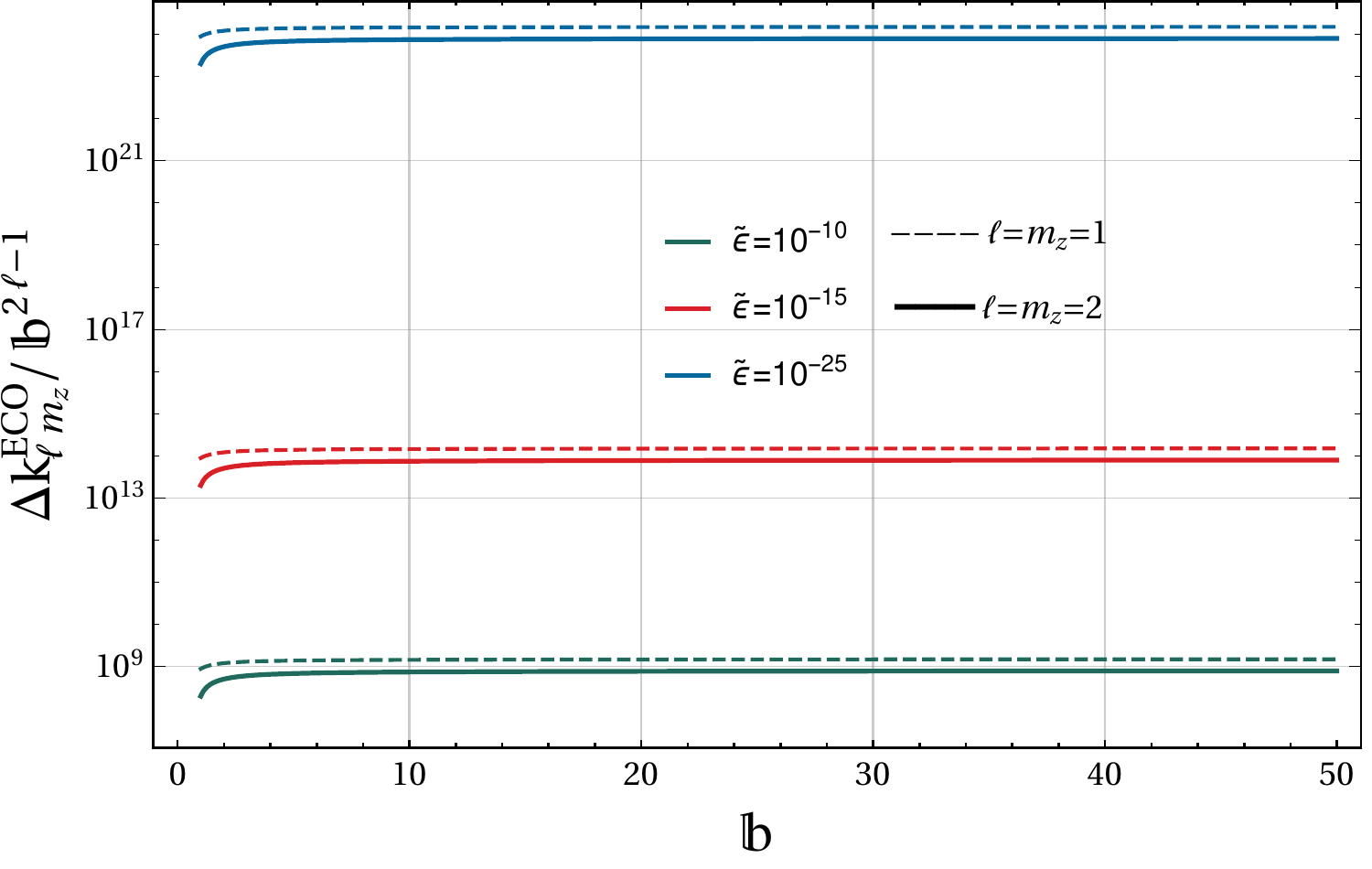}
\caption{Here we depict the quantity $(\Delta k_{\ell~m_z}^{\rm ECO}/\mathbb{b}^{2\ell-1})=\lt|\lt(k_{l~m_z}^{\rm ECO+Acc}/k_{l~m_z}^{\rm ECO}\rt)-1\rt|\mathbb{b}^{-2\ell+1}$, shown as a function of $\mathbb{b}$, for $\ell=1=m_z$ and $\ell=2=m_z$ at fixed $\tilde{\epsilon}$ with $\epsilon=0.01$ and $\mathcal{R}=-1$.}
\label{fig:LOmega4}}
\end{figure}

To elaborate the dependence of the static LNs for accreting Schwarzschild-like ECOs on the scale $\mathbb{b}$, we first define the fractional change in the LNs of an accreting ECO due to the accretion disk as, $\Delta k_{\ell m_z}^{\rm ECO}\equiv \lt|\lt(k_{\ell m_z}^{\rm ECO+Acc}/k_{\ell m_z}^{\rm ECO}\rt)-1\rt|$. Then we plot the ratio $(\Delta k_{\ell m_{z}}/b^{2\ell-1})$ with the dimensionless scale factor $\mathbb{b}$ for $j_t=0$ in \cref{fig:LOmega3} for an ECO satisfying Dirichlet boundary condition ($\mathcal{R}=1$), and similarly in \cref{fig:LOmega4} for Neumann boundary condition ($\mathcal{R}=-1$).

From \cref{fig:LOmega3} and \cref{fig:LOmega4}, we observe that in the large-$\mathbb{b}$ regime, the fractional change in the static LNs of an ECO due to accretion, namely $\Delta k_{\ell m_z}^{\rm ECO}$, exhibits a clear power-law behaviour with $\mathbb{b}$, scaling as $\mathbb{b}^{3}$ for $\ell=2=m_{z}$ and as $\mathbb{b}$ for $\ell=1=m_{z}$ (and hence as $\mathbb{b}^{2\ell-1}$ for generic $\ell=m_{z}$). This behaviour indicates that the influence of accretion on the static LNs becomes increasingly significant as $\mathbb{b}$ grows, resulting in a rapid enhancement of the deviation relative to the non-accreting case. This growth of LNs with increasing $\mathbb{b}$ for a fixed disk mass can be attributed to the dilution of the accretion disk. As the same mass is now distributed over a larger length scale, the disk becomes more dilute and hence can be tidally distorted more easily, leading to larger deformability and hence larger LNs. Note that the above scaling of the accretion-induced LNs with $\mathbb{b}$ is also coincident with the LNs of BH in the presence of an accretion disk, see \cref{eq:LoveBHacc-large}.

So far, all the results were presented for the $j_{t}=0$ case. While for $j_t>0$, the contribution of the accretion disk on spacetime geometry arises through the metric functions $\nu_{\rm disk}$ and $\lambda_{\rm int}$, both of which become explicit functions of $\theta$, see \cref{eq:nu-lamda}, and the Logarithmic term in \cref{eq:ansatz} becomes non vanishing. This results in the static LNs developing Logarithmic dependence on the radial coordinate $r$, namely,
\begin{align}\label{eq:tab}
k_{\ell m_z}^{\rm ECO+Acc}= A_{\ell m_z}+B_{\ell m_z}\log(r/2M)+\mathcal{O}\lt(\epsilon\tilde\epsilon, {\tilde\epsilon}^2\rt)~.
\end{align}
We compute a list of representative values of $A_{\ell m_{z}}$ and $B_{\ell m_{z}}$ for different values of $j_t$ and for reflectivities $\mathcal{R}=\pm 1$ with $\mathbb{b}=12$, $\epsilon=0.01$ and $\widetilde\epsilon=10^{-8}$ in Table \ref{table:jt}.

\begin{table}[htbp]
\begin{center}
\begin{ruledtabular}
\begin{tabular}{cccccc}
$j_t$ & $\mathcal{R}$ & $A_{22}$ & $B_{22}$& $A_{11}$ & $B_{11}$ \\ 
\hline
\multirow{2}{*}{1}
  & { }1  & -21.00 & 12.14& 0.03 & -0.02 \\
  & -1  & -20.79 & 12.02 & 0.04 & -0.02\\
\hline
\multirow{2}{*}{2}
  &  { }1  & -43.23 & 12.14 & 0.06 & -0.02\\
  & -1  & -42.79 & 12.02 & 0.07 & -0.02\\
\hline
\multirow{2}{*}{4}
  &  { }1  &  12.74 & -1.35 &  0.02 & -0.01\\
  & -1  &  12.62 & -1.34&  0.02 & -0.01 \\
\hline
\multirow{2}{*}{6}
  &  { }1  & -7.00 & 3.37 &  0.02 & -0.01\\
  & -1  & -6.93 & 3.34 &  0.02 & -0.01\\
\hline
\multirow{2}{*}{8}
  &  { }1  & -6.64 & 3.37 &  0.02 & -0.01\\
  & -1  & -6.57 & 3.34 &  0.02 & -0.01\\
\hline
\multirow{2}{*}{10}
  &  { }1  & -6.58 & 3.37 &  0.02 & -0.01\\
  & -1  & -6.52 & 3.34&  0.02 & -0.01 \\
 \end{tabular}
\end{ruledtabular}
\caption{Representative numerical values for $A_{\ell m_{z}}$ and  $B_{\ell m_{z}}$ for different values of $j_t$ and $\mathcal{R}=\pm 1$ with $r_{\rm h}=1$, $\mathbb{b}=12$, $\epsilon=0.01$, and compactness of the ECO being $\widetilde \epsilon=10^{-8}$.}
\label{table:jt}
\end{center}
\end{table}

\begin{figure}[!htbp]
\centering
{\includegraphics[width=\linewidth]{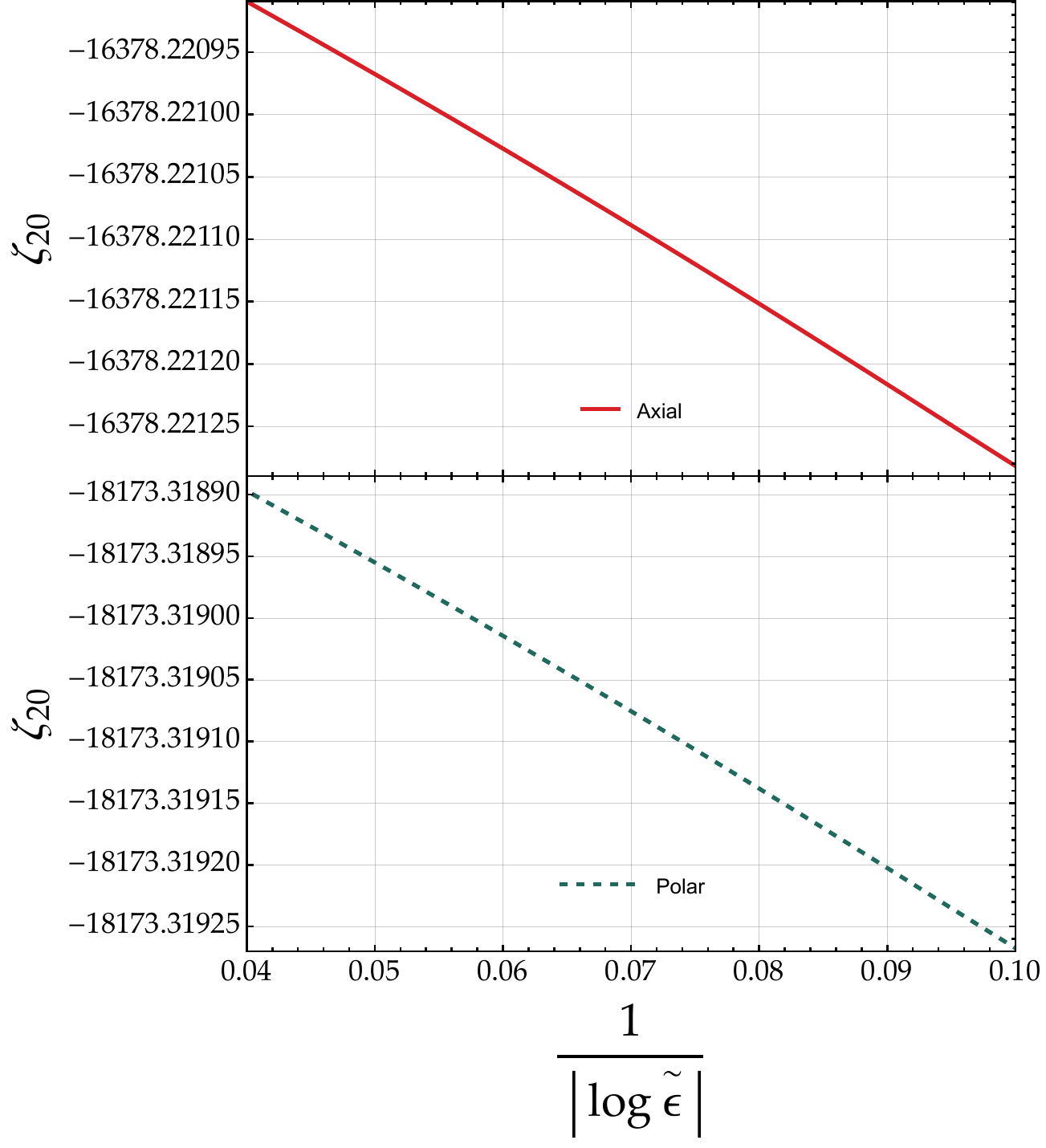}}
\caption{Variation of $\zeta_{20}$ in the axial (top panel) and polar sector (bottom pannel) for spin-1 perturbation of an accreting ECO, with $\mathcal{R}=1$, plotted with $\lt|\log\tilde{\epsilon}\rt|^{-1}$ for the $\ell=2$, $m_{z}=0$ mode. For these plots, we have chosen $\epsilon=0.01$ and $\mathbb{b}=20$.} 
\label{fig:Loveepsilon}
\end{figure}
\begin{figure}[!htbp]
\centering
{\includegraphics[width=0.95\linewidth]{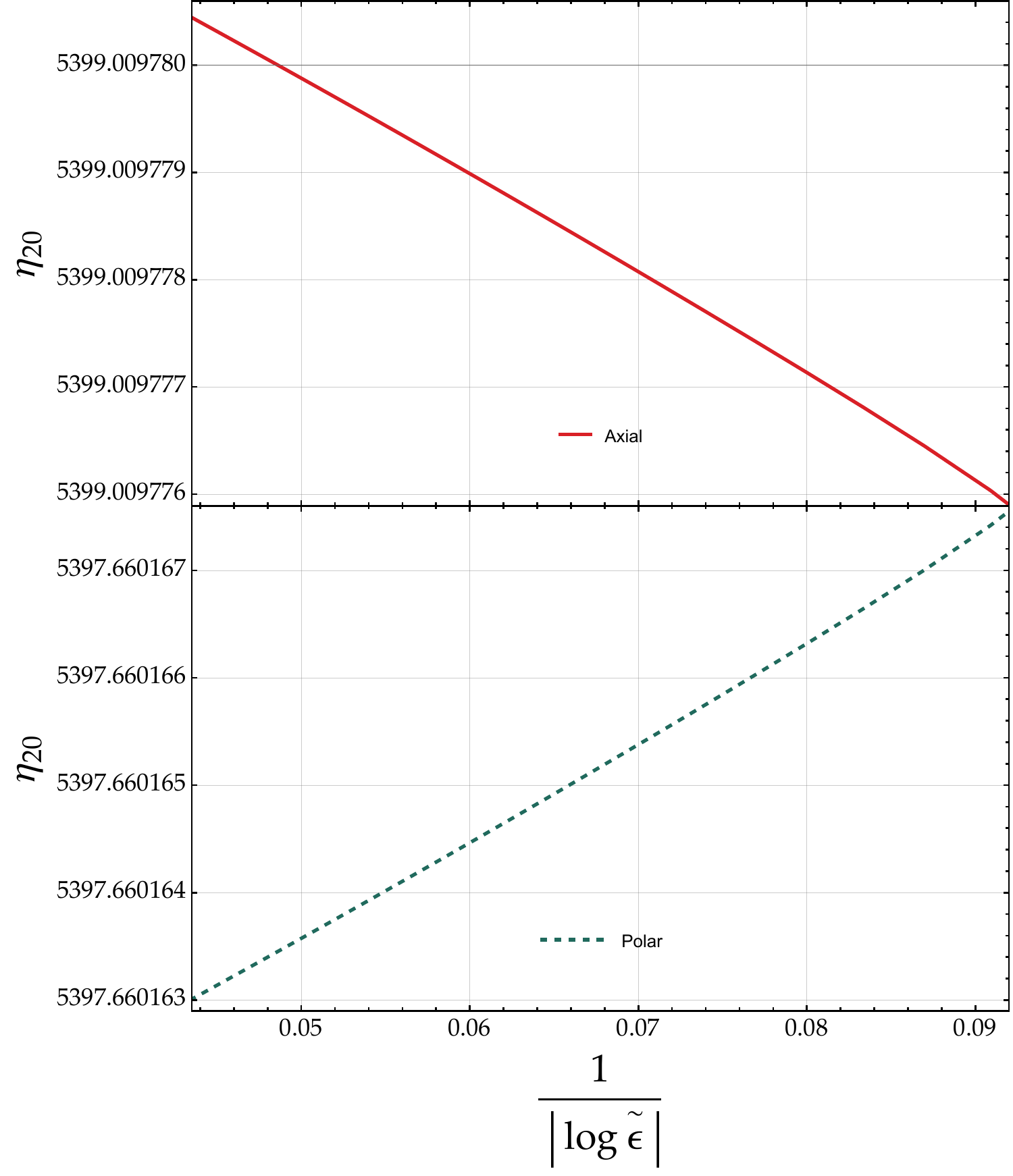}}
\caption{Variation of $\eta_{20}$ in the axial (top panel) and polar sector (bottom pannel) for spin-1 perturbation of an accreting ECO with $\mathcal{R}=1$, plotted with $\lt|\log\tilde{\epsilon}\rt|^{-1}$ for the $\ell=2$, $m_{z}=0$ mode. Here also, the accretion related parameters are chosen to be $\epsilon=0.01$ and $\mathbb{b}=20$.} 
\label{fig:Loveepsilon-eta}
\end{figure}
\begin{figure}[!htbp]
\centering
{\includegraphics[width=0.95\linewidth]{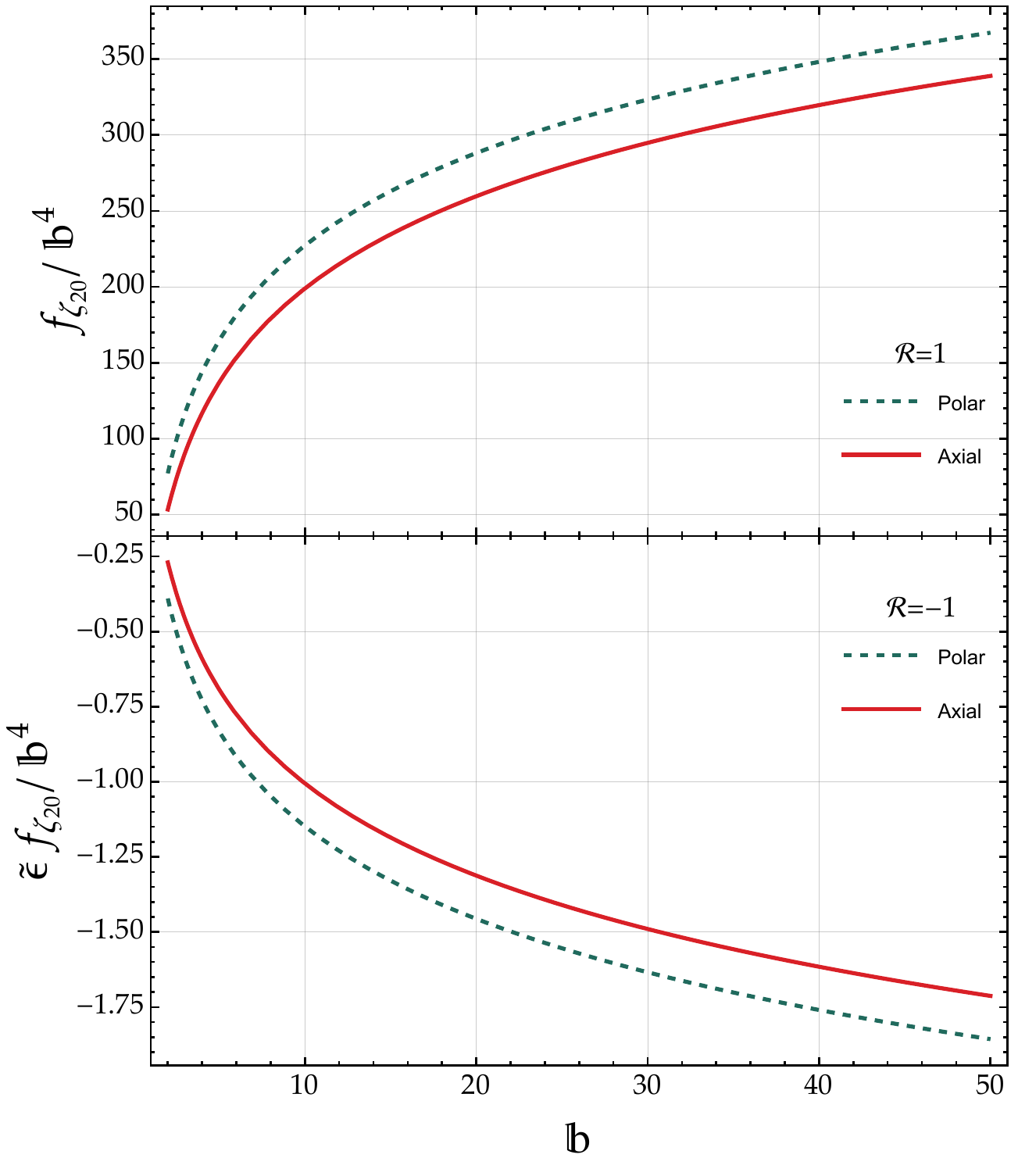}
\caption{The top figure displays the scaling of the quantity $f_{\zeta_{20}}\equiv
(\zeta_{20}/k_{20}^{\rm ECO})$, for an accreting ECO with $\mathcal{R}=1$, with the accretion scale $\mathbb{b}$. While the bottom figure shows the scaling of $\widetilde{\epsilon}f_{\zeta_{20}}$ for an accreting ECO with $\mathcal{R}=-1$, with the accretion scale $\mathbb{b}$. Both the plots are for spin-1 perturbation at fixed $\tilde{\epsilon}=10^{-15}$ and $\epsilon=0.01$.}
\label{fig:LOmegaEM1}}
\end{figure}
\begin{figure}[!t]
\centering
{\includegraphics[width=\linewidth]{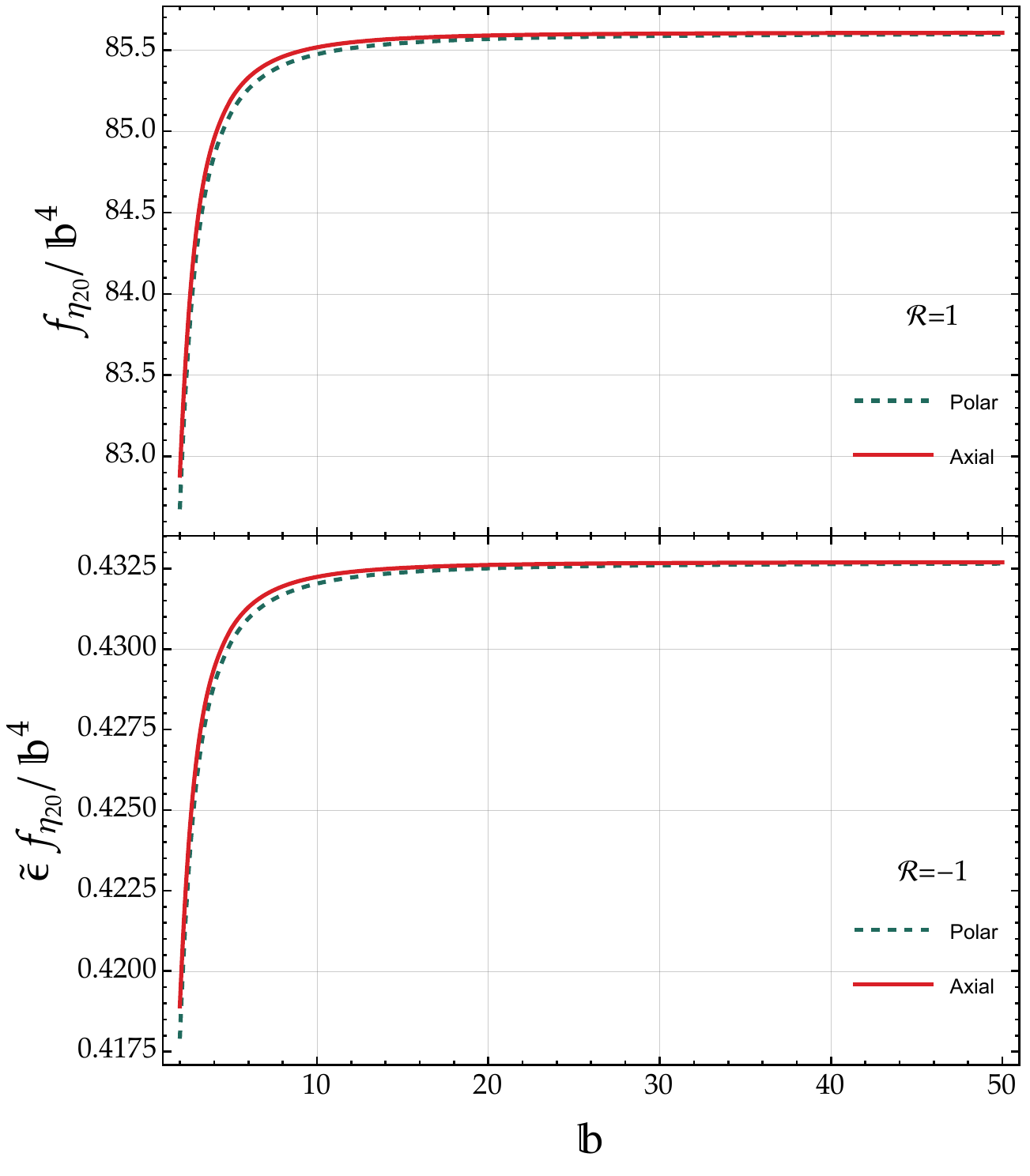}
\caption{The top figure displays the scaling of the quantity $f_{\eta_{20}}\equiv (\eta_{20}/k_{20}^{\rm ECO})$ with the accretion scale $\mathbb{b}$, for an accreting ECO with $\mathcal{R}=1$. The bottom figure, on the other hand, describes the scaling of $\widetilde{\epsilon}f_{\eta_{20}}$ with $\mathbb{b}$ for an accreting ECO with $\mathcal{R}=-1$. For both the plots, we have fixed $\tilde{\epsilon}=10^{-15}$ and $\epsilon=0.01$ for the spin-1 perturbation considered here.}
\label{fig:LOmegaEM2}}
\end{figure}

\subsection{Spin-1 sector}

Analogous to the case of scalar LNs for an accreting ECO, the static LNs of an accreting ECO under a spin-1 perturbation, in both the polar and axial sectors, exhibit a logarithmic dependence on the radial coordinate $r$, namely,
\begin{align}\label{eq:tab1}
k_{\ell m_z}^{\rm ECO+Acc}={\zeta}_{\ell m_z}+\eta_{\ell m_z}\log\left(\frac{r}{r_h}\right)\,.
\end{align}
Note that the above Logarithmic dependence holds for generic choices of $j_{t}$, including $j_{t}=0$. In what follows, we specialize to the case of $\ell=2$ and $m_{z}=0$ for simplicity. However, all the results can be straightforwardly generalized for the case of $m_{z}=\pm 1$ and $m_{z}=\pm 2$, as well.  

We have plotted $\zeta_{20}$ and $\eta_{20}$ with $\lt|\log{\widetilde{\epsilon}}\rt|^{-1}$ for $\epsilon=0.01$ and $\mathbb{b}=20$ in \cref{fig:Loveepsilon,fig:Loveepsilon-eta}, respectively. We note that for sufficiently compact (small values $\widetilde{\epsilon}$) and perfectly reflecting ($\mathcal{R}=1$) ECOs, both $\zeta_{20}$ and $\eta_{20}$ are linear in $\lt|\log\widetilde{\epsilon}\rt|^{-1}$. Thus, the characteristic logarithmic sensitivity of static LNs on the ECO compactness, see~\cref{eq:TLN-spin1-eco}, persists even in the presence of an accretion disk for the spin-1 perturbation in both the polar and axial sectors. For $\mathcal{R}=1$, the top figures of \cref{fig:LOmegaEM1} and \cref{fig:LOmegaEM2} show the scaling of the ratio, $f_{\zeta_{20}}=(\zeta_{20}/k_{20}^{\rm ECO})$ and $f_{\eta_{20}}=
(\eta_{20}/k_{20}^{\rm ECO})$ with $\mathbb{b}$ at fixed $\epsilon$ and $\widetilde{\epsilon}$, for the  spin-1 case. For $\mathcal{R}=-1$, on the other hand, $k_{20}^{\rm ECO}$ scales as $\widetilde{\epsilon}$, and hence we have depicted the scaling of $\widetilde{\epsilon}f_{\zeta_{20}}$ and $\widetilde{\epsilon}f_{\eta_{20}}$ with $\mathbb{b}$ in \Cref{fig:LOmegaEM1} and \cref{fig:LOmegaEM2}, respectively. It is evident that in the large-$\mathbb{b}$ regime, $\eta_{20}$ exhibits a clear power-law behaviour, scaling as $\mathbb{b}^4$, although $\zeta_{20}$ shows a non-trivial dependence on $\mathbb{b}$, over and above the $\mathbb{b}^{4}$ behaviour, for both $\mathcal{R}=\pm 1$.

\subsection{Intrinsic versus accretion-induced response}\label{sec:new}

In this section, we would like to discuss the possibilities to distinguish between the intrinsic and accretion-induced responses of an ECO due to an external tidal field and possible observational implications of the same. For simplicity, we concentrate on the $j_t=0$ case and scalar perturbations, for which the source term appearing in \cref{split} has the following structure: $\mathbb{V}=m_z(2\ell+1)\mathcal{L}_0$. One can easily verify from \cref{eq:L0} that the quantity $\mathcal{L}_0$ has a dominant contribution from $r\sim b\gg M$, since that is where most of the disk mass resides. Thus, taking appropriate limits of \cref{eq:L0}, we obtain the following behaviour for the source term appearing in the expression for static LNs, 
\begin{align}\label{eq:Vscal}
\mathbb{V}\simeq-m_{z}(2\ell+1)\,\frac{M^{2}r}{(b^{2}+r^{2})^{3/2}}\,.
\end{align}
For separating the intrinsic tidal LNs of an ECO from the accretion-induced tidal effects, we start with \cref{split}, which explicitly demonstrates that the LN of an accreting ECO, at leading order in $\epsilon$, can be expressed as the sum of the intrinsic LN $k^{\rm ECO}_{\ell m_z}$, the LN $k^{\rm BH+Acc}_{\ell m_z}$ of an accreting BH with the same accretion disk, and an additional term $\alpha_{\ell m_{z}}^{\rm extra}k^{\rm ECO}_{\ell m_z}$. Thus accretion-induced LNs for an ECO have a part $k^{\rm BH+Acc}_{\ell m_z}$, which does not depend on the boundary condition at the ECO surface and is solely generated by accretion, while $\alpha_{\ell m_{z}}^{\rm extra}$ encodes the effect of the reflective boundary condition on the accretion-induced LNs for an ECO.

We start with the pure accretion-induced term in the LN, $k^{\rm BH+Acc}_{\ell m_z}$, which is independent of the nature of the central compact object. By \cref{LoveBHGen}, the dependence of this term on the accretion scale $\mathbb{b}$ is set by the following integral: $\int_{r_0}^{b}P_\ell\,\mathbb{V}\,P_\ell\,dr'$. Since, the dominant contribution to this integral comes from $r\sim b\gg M$, we can use the asymptotic form for the Legendre polynomial, such that, $P_{\ell}^{2}\simeq p_\ell^2(r/2M)^{2\ell}$. Hence the above integral boils down to,
\begin{align}\label{eq:kBHlarge}
k^{\rm BH+Acc}_{\ell m_z}&=\frac{\epsilon Mm_{z}(2\ell+1)q_{\ell}p_{\ell}}{2(2M)^{2\ell}}
\int_{r_{0}}^{b}dr'\frac{r'^{2\ell+1}}{(b^{2}+r'^{2})^{3/2}}\,,
\nonumber
\\
&\simeq\frac{\epsilon m_z}{8}\left(\frac{\mathbb{b}}{2}\right)^{2\ell-1}+\mathcal{O}(\mathbb{b}^{2\ell-2})\,,
\end{align}
where, we have used the result $p_\ell q_\ell=\{1/2(2\ell+1)\}$, and have ignored numerical factors in the final expression, arising from the integral. Thus for $\ell=2=m_{z}$, we obtain, $k^{\rm BH+Acc}_{22}\simeq\epsilon\,\mathbb{b}^3$, which has the same scaling with $\mathbb{b}$, as the leading order term of \cref{eq:LoveBHacc-large}. Similarly, for $\ell=1=m_{z}$, we obtain $k^{\rm BH+Acc}_{11}\simeq\epsilon\,\mathbb{b}$. Thus, the accretion-induced LN, which is independent of the nature of the compact object grows with the accretion length scale as $\mathbb{b}^{2\ell-1}$. 

The accretion-induced, but compact object dependent term, on the other hand, is given by $\alpha_{\ell m_{z}}^{\rm extra}$, see \cref{split}, and involves integrals of $Q_{\ell}^{2}\mathbb{V}$ and $P_{\ell}Q_{\ell}\mathbb{V}$. For, $r\sim b\gg M$, we can use the asymptotic expansion for the Legendre polynomial of the second kind $Q_{\ell}$ as well, such that $\alpha_{\ell m_{z}}^{\rm extra}$ reduces to,
\begin{align}\label{eq:Ilarge}
&\alpha^{\rm extra}_{\ell m_z}=2\epsilon m_{z}M^{2}\int_{r_{0}}^{b}\frac{dr'}{(b^{2}+r'^{2})^{3/2}}\,
\nonumber
\\
&+2\epsilon m_{z}M^{2}(2M)^{2\ell+1}k_{\ell m_{z}}^{\rm ECO}\int_{r_{0}}^{b}\frac{dr'r'^{-2\ell-1}}{(b^{2}+r'^{2})^{3/2}}
\\
&\simeq\frac{2\epsilon m_z}{\mathbb{b}^2}+\mathcal{O}(\mathbb{b}^{-3})\,.
\end{align}
where, we have used the result that $p_\ell q_\ell=\{1/2(2\ell+1)\}$. Thus, there is a clear separation of scales in the expression for the LNs, the accretion-induced term, independent of the nature of the compact object, grows as $\epsilon \mathbb{b}^{2\ell-1}$, while the accretion-induced, but compact object dependent terms are of order $\epsilon\,k^{\rm ECO}_{\ell m_z}\mathbb{b}^{-2}$ or smaller. Therefore, we obtain the following scaling of the LNs for an accreting ECO with the scale $\mathbb{b}$ of the accretion disk, 
\begin{align}\label{eq:finalLNeco}
k_{\ell m_{z}}^{\rm ECO+Acc}=\left(1+\frac{2\epsilon m_{z}}{\mathbb{b}^{2}}\right)k_{\ell m_{z}}^{\rm ECO}+\frac{\epsilon m_z}{8}\left(\frac{\mathbb{b}}{2}\right)^{2\ell-1}\,.
\end{align}
In what follows, we will verify the above result explicitly for the $\ell=2=m_{z}$ and $\ell=1=m_{z}$ mode, using the numerical procedure laid out in \cref{sec:eco-dsk-tln}. 

\begin{figure}[!htbp]
\centering
\includegraphics[width=\linewidth]{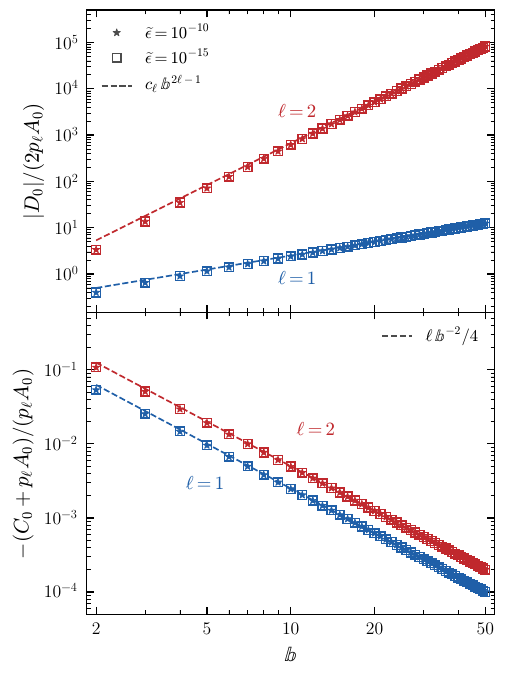}
\caption{Scaling of the appropriately normalised coefficients of the far-zone ansatz~\eqref{eq:ansatz} with $\mathbb{b}$ has been presented for $\ell=2=m_z$ (in red) and $\ell=1=m_z$ (in blue), with $j_t=0$, $\mathcal{R}=1$, and $\widetilde{\epsilon}=10^{-10}$ (stars) and $10^{-15}$ (squares). Top: the plot depicts the matching between the ratio $|D_0|/(2p_\ell A_0)$, represented by points, with $(k^{\rm BH+Acc}_{\ell m_z}/\epsilon)$, depicted by dashed lines, and scaling as $\mathbb{b}^{2\ell-1}$. Bottom: the coincident scaling of the plot of the points depicting the following combination $-(C_0+p_\ell A_0)/(p_\ell A_0)$, along with accretion-induced, compact object dependent LN, presented with dashed lines, and scaling as $\mathbb{b}^{-2}$.}
\label{fig:C0D0}
\end{figure}

The starting point of the numerical procedure is \cref{eq:ansatz}, which we fit to the numerical solution of \cref{eq:psi1}, for the linear-in-$\epsilon$ perturbation variable $\Psi^{(1)}_{\ell m_{z}}$. We perform the fitting for both $\ell=2=m_{z}$ and $\ell=1=m_{z}$ modes, with two different choices of $\widetilde{\epsilon}$ and for several different accretion length scales $\mathbb{b}$. The result of the analysis has been presented in \cref{fig:C0D0}.

As evident, both for $\ell=2=m_{z}$ and for $\ell=1=m_{z}$ modes, the LNs of an accreting ECO are mostly governed by their scaling with $\mathbb{b}$, while the dependence on the compactness $\widetilde{\epsilon}$ is minimal. Further, \cref{fig:C0D0} depicts that the coefficient $(D_{0}/A_{0})$ in \cref{eq:ansatz} scales as $\mathbb{b}^{2\ell-1}$, which is consistent with our previous analytical findings, as this ratio is nothing but the accretion-induced LN, which is independent of the nature of the compact object, see \cref{eq:kBHD0}. Similarly, as \cref{fig:C0D0} demonstrates, the ratio $(C_{0}/A_{0})$, appearing in \cref{eq:ansatz} scales as $\mathbb{b}^{-2}$, which from our analytical treatment is precisely the scaling of accretion-induced, compact object dependent LN, see \cref{eq:c0d0}. Thus, the numerical analysis of the perturbation equation is also consistent with our analytical result, presented in \cref{eq:finalLNeco}. 

Having described the consistency of the analytical result, presented in \cref{eq:finalLNeco}, with the numerical evaluation of the static LNs of an accreting ECO, let us comment on the issue of degeneracy. First of all, since the typical size of the accretion disk is much larger than the size of the compact object ($\mathbb{b}\gg 1$) the presence of any accretion disk will enhance the LNs by a factor of $\mathbb{b}^{2\ell-1}$, while at the same time will be reduced by a factor of $\epsilon$. In particular, for $\ell=2=m_z$ and $\mathcal{R}=1$, if $\epsilon=$(disk mass/BH mass)$>\{240\,\mathbb{b}^3|3+\log\widetilde{\epsilon}|\}^{-1}$, it follows that the contribution from the accretion disk always dominates. Of course, the nature of the compact object is always imprinted on the LNs as well, but their magnitudes are much suppressed and have the universal Logarithmic scaling with the compactness. For BHs, on the other hand, there is no such Logarithmic scaling. 

Therefore, for very compact ($\mathbb{b}\geq 1$) and small accretion disk $\epsilon<\epsilon_{\rm c} \ll 1$, it is indeed possible to decipher the nature of the central compact object, as the contribution from the ECO dominates over that of the accretion disk. However, for typical accretion disks, with $\mathbb{b}\gg 1$ and $\epsilon >\epsilon_{c}$, the contribution from the accretion disk dominates the LNs, and hence, it is very difficult to distinguish between accreting BHs and accreting ECOs through tidal observations alone. Of course, by combining more than one observable, it is entirely possible to decipher the nature of the central compact object, e.g., the measurement of LNs will give information about the accretion disk, while the measurement of quasi-normal modes will provide a tell-tale signature of the nature of the central compact object.

Finally, we would like to point out a few other interesting features of our result. (a) Axisymmetric perturbations ($m_{z=0}$) do not feel the presence of the disk, and hence at linear order in $\epsilon$, we have $k^{\rm ECO+Acc}_{\ell0}=k^{\rm ECO}_\ell$. Thus by comparing LNs corresponding to axisymmetric and non-axisymmetric modes, we can distinguish between accreting BH and accreting ECO. This does not hold for the spin-1 perturbation. (b) The contributions to the LNs from accretion and from the central compact object, scale differently with $\ell$. Thus, measurement of several multipolar LNs can therefore separate between accretion-induced and intrinsic LNs of the compact object. (c) For $j_t>0$ there is a logarithmic running of the accretion-induced LNs, while the ECO term exhibits no such Logarithmic running. (d) Finally, in the context of extreme mass ratio inspiral, the disk affects the orbit of the secondary, through the modified spacetime geometry, as well as through dynamical friction \cite{Radulea:2026hub}, both of which appears at a lower post-Newtonian order than the tidal effects, and hence can be used to break the degeneracy.

\section{Summary and Discussion}\label{sec:summary}

Tidal deformability of compact objects is crucial for understanding not only the objects themselves but also their surroundings and the underlying laws of gravity that govern their dynamics. While isolated BHs in four dimensional general relativity are special as they have vanishing static LNs under bosonic perturbation, the presence of an accretion disk/dark matter halo can endow them with non-zero LNs~\cite{Cannizzaro:2024fpz, Chakraborty:2024gcr,Bhowmik:2026owi,Ghosh:2026fbr}. Since ECOs help in removing the horizon of a BH, and thereby excising the existence of a singularity, non-deterministic evolution due to the Cauchy horizon and the information loss paradox, it is important to answer the following related questions: (i) how does the presence of an accretion disk affect the LNs of ECOs, which otherwise have non-zero LNs? (the case of an ECO within a dark matter halo has already been presented in \cite{Chakravarti:2025awj}) (ii) Can the presence of an accretion disk mask the near-horizon modifications in the case of an ECO? As a first step towards addressing these questions, we performed a detailed analysis of the static scalar and spin-1 LNs of accreting spherically symmetric ECOs in the present work. 

We modelled the spacetime as that of a central, spherically symmetric, compact object surrounded by a thin axi-symmetric accretion disk. In the absence of accretion, the geometry outside the ECO is given by the Schwarzschild spacetime, while the accretion disk itself is considered to be of finite mass and of infinite spatial extent, modifying the spacetime geometry and extending from the surface of the object and localized on the equatorial plane. We assumed the mass $\mathcal{M_{\rm d}}$ of the accretion disk to be much smaller than the mass $M$ of the compact object $(\epsilon\equiv \mathcal{M}_{\rm d}/M\ll1)$. Constrained to the leading order in $\epsilon$, we note that the exterior geometry can be well described by a deformed Schwarzschild metric, see~\cref{eq:gtt,eq:grr,eq:gthth,eq:gphph}. Further, the axi-symmetric nature of the accretion disk breaks the spherical symmetry of the background spacetime and hence the perturbations depend on the azimuthal number $m_z$ as well, besides the angular number $\ell$. This shows that the perturbation equations of a spherically symmetric compact object living in an accreting environment are \emph{not} spherically symmetric. The properties of the ECOs, on the other hand, are essentially encoded by two quantities: (i) the compactness $(M/r_0)$ of the ECO, where $r_{0}=2M(1+\widetilde{\epsilon})$ is the radius of the ECO and (ii) the surface reflectivity $\mathcal{R}$. Note that for an ECO of a fixed mass $M$, the compactness can be equivalently characterised by the parameter $\widetilde{\epsilon}$ defining the location of its surface.

Given this setup, our results demonstrate that the LNs of a perfectly reflecting ECO, in the presence of an accretion disk, retain the logarithmic dependence on the compactness parameter $\widetilde{\epsilon}$ for both scalar and spin-1 perturbations (see, \cref{fig:LOmega1}, \cref{fig:Loveepsilon} and \cref{fig:Loveepsilon-eta}). This logarithmic scaling of the LNs is governed primarily by the near-horizon structure of the ECO and remains qualitatively unaltered by the presence of accreting matter. Accretion, therefore, does not modify the underlying scaling behaviour associated with the compactness of the object, but instead affects the overall magnitude of the static response, through an additive accretion disk induced contribution. 

At leading order in $\epsilon$, the static response is the sum of the intrinsic response of the ECO, the response of an accreting BH with the same disk, and the response of the accretion disk with a modified ECO boundary condition, see \cref{eq:finalLNeco}. The response of the accretion disk with modified ECO boundary condition is suppressed by $\mathbb{b}^{-2}$, where $\mathbb{b}\equiv (b/M)$, with $b$ being a characteristic scale of the accretion disk. The contribution from the accretion disk alone grows rapidly with the size of the disk, as $\mathbb{b}^{2\ell-1}$. This behaviour has been numerically verified and has been depicted in \cref{fig:LOmega3,fig:LOmega4,fig:C0D0}. For spin-1 perturbations, on the other hand, the LNs experience a Logarithmic running, as $k_{\ell m}\sim\zeta_{\ell m}+\eta_{\ell m}\log(r/2M)$. For $\ell=2$ and $m_{z=0}$ mode, the coefficient $\eta_{20}$ grows as $\mathbb{b}^{4}$, while $\zeta_{20}$ has a more involved dependence on $\mathbb{b}$ (see \cref{fig:LOmegaEM1,fig:LOmegaEM2}). Thus, irrespective of the nature of the perturbation, the accretion disk induced LNs grow as the size of the disk becomes bigger. This growth reflects the larger deformability of a more diluted mass distribution.

These results bear directly on the observability of the LNs for accreting compact objects. For both BHs and ECOs, the LNs are either zero, or small (due to the dependence of the LNs on the compactness only through a logarithm), but the accretion induced effects are significantly larger. This suggests that detecting the tidal deformability of an accreting compact object is much easier than that of the compact object itself. In fact, only when the disk mass is lower than a small fraction of the mass of the central compact object, about $10^{-6}$ for realistic $\mathbb{b}$, the intrinsic response of a perfectly reflecting ECO becomes sufficiently large, making its distinction from that of a BH possible. 

Thus, for a substantial region of the parameter space of $\mathbb{b}$, the measurement of the static LN for a given $(\ell,m_{z})$ mode cannot distinguish an accreting ECO from an accreting BH. This degeneracy can be broken in principle through the measurement of static LNs for different multipolar modes, e.g., the axisymmetric modes have vanishing contribution from the accretion disk, and if found to be non-zero, is a tell-tale signature of the ECOs. Similarly, in generic cases one expects a Logarithmic running of the LNs, which is a purely accretion-induced effect and has no analog for isolated ECOs. However, a similar degeneracy between accreting and a dark-matter environment is much harder to distinguish.

The present analysis can be extended in several directions. First, the rotation of the central compact object has been neglected. Incorporating spin would require moving to stationary axisymmetric spacetimes that include frame-dragging effects, thereby modifying both the disk structure and the perturbation sector. Second, while scalar and spin-1 perturbations can provide qualitative insight into environmental effects, they do not capture the full metric–matter coupling present in gravitational perturbations. A more complete treatment of tidal deformations in binary inspiral scenarios would therefore require a dedicated analysis of gravitational perturbations in this background. Such an analysis would also show whether the additive structure of the LNs found here, and the suppression of the accretion-induced term, which are dependent on boundary effects, carry over to the gravitational LNs as well. Third, the assumption of stationarity limits the applicability of the present model for describing electromagnetic emission from realistic accretion flows. The disk model adopted here does not include angular momentum transport, viscous dissipation, or radiative processes, and should therefore be regarded as an effective description of a quasi-stationary matter distribution rather than a fully self-consistent accretion model. In particular, it would be interesting to address issues involving possible accumulation of matter near the compact-object surface, and the associated formation of shocks, that could modify both the disk structure and the associated LNs. From a multi-messenger perspective, particularly in binary coalescence scenarios, the coupled evolution of the disk, spacetime, and compact-object surface may become dynamically important. A fully self-consistent treatment would likely require time-dependent modelling, potentially requiring more robust numerical approaches.

\section*{Acknowledgment}

We thank Valerio De Luca and Enrico Cannizzaro for their useful discussions. We acknowledge hospitality at ICTS Bangalore during the program - ``Beyond the Horizon: Testing the black hole paradigm'' (code: ICTS/BTH2025/03), and IIT Gandhinagar during the program ``GW10'', where a part of the work was done. AC and CS also extend their gratitude to IACS Kolkata for hospitality and computational support. AC also acknowledges hospitality at IUCAA Pune. AC thanks G. Sen for discussions during the early phase of the work. The work of AC was partly funded by the National Postdoctoral Fellowship of the Anusandhan National Research Foundation (ANRF), Govt. of India (File No.: PDF/2023/000550) at IIT Guwahati. The work of Kazuharu Bamba was supported in part by the JSPS KAKENHI Grants No.~24KF0100
and No.~25KF0176, and a grant-in-aid of academic research of the Yamaguchi Scholarship Foundation. The research of SC is supported by MATRICS (MTR/2023/000049) and Core Research Grants (CRG/2023/000934) from SERB, ANRF, Government of India.  SC also thanks the local hospitality at ICTS and IUCAA through the associateship program, where a part of this work was done. 

\appendix
\section{Thin Disk Solutions in Weyl Coordinates}\label{metric}

\subsection{Kuzmin-Toomre disks}

In static and axisymmetric spacetimes of the Weyl class, see \cref{sec:BG} for details, the metric function $\nu$ satisfies the Laplace's equation in vacuum and may therefore be generated from known Newtonian potentials. Since we are interested in thin accretion disk models, a particularly useful class of seed solutions is provided by infinitesimally thin disks, whose gravitational fields can be constructed analytically through the displace-cut-reflect method.

The simplest example is the Kuzmin disk \cite{Kuzmin}. It is obtained by placing a point mass $\mathcal{M}_{\rm d}$ at a distance $b$ below the equatorial plane and replacing $z\rightarrow \lt|z\rt|$ in the gravitational potential obtained at $z>0$. The resulting potential,
\begin{equation}
\nu^{\rm (disk)}_{\rm K}=-\frac{\mathcal{M}_{\rm d}}{\sqrt{\rho^2+(|z|+b)^2}}\,,
\end{equation}
satisfies Laplace's equation everywhere except at $z=0$, where the discontinuity in the normal derivative at $z=0$ gives rise to a razor-thin disk source. Thus, even though we started with a monopole displaced from the origin, the resulting potential with $z\to \lt|z\rt|$ transformation leads to the following surface density,
\begin{equation}
\Sigma_{\rm K}=\frac{\mathcal{M}_{\rm d} b}{2\pi(\rho^2+b^2)^{3/2}}\,,
\end{equation}
and describes an infinitely extended disk with finite total mass. The density is maximal at the center and decreases as $\Sigma_{\rm K}\sim \rho^{-3}$ at large radii.

The gravitational potential $\nu_{\rm K}^{\rm (disk)}$ of the Kuzmin disk is not a unique solution to Poisson's equation with source confined to the $z=0$ plane, in cylindrical coordinates. There can be infinitely many choices. The most general disk model of this kind, known as the Kuzmin-Toomre family \cite{1963ApJ...138..385T}, is sourced by a matter distribution strictly on the $z=0$ plane, and involves the following potential
\begin{equation}
\nu^{\rm (disk)}_{n\rm (KT)}=-\frac{\mathcal{M}_{\rm d}}{(2n-1)!!}\sum_{k=0}^{n}\frac{(2n-k)!}{2^{\,n-k}(n-k)!}\frac{b^k}{r_b^{k+1}}P_{k}\left(\frac{|z|+b}{r_{\rm b}}\right)\,.
\end{equation}
Here, the quantity $r_{\rm b}$ has been defined in \cref{definition_rb} in the main text and $P_{k}(x)$ is the Legendre polynomial of order $k$. The integer $n$ labels different disk profiles within the same family. The original Kuzmin disk corresponds to the lowest-order member of the family ($n=0$), while higher values of $n$ generate progressively steeper fall off for the density profiles.

The corresponding Newtonian surface density is
\begin{equation}
\Sigma_{n\rm (KT)}=\frac{(2n+1)b^{2n+1}}{2\pi}\frac{\mathcal{M}_{\rm d}}{(\rho^2+b^2)^{n+3/2}}\,,
\end{equation}
with the following asymptotic behaviour,
\begin{equation}
\Sigma_{n\rm{(KT)}}\sim\rho^{-2n-3}\,,
\qquad
\rho\rightarrow\infty\,.
\end{equation}
Thus, increasing $n$ concentrates a larger fraction of the mass toward the inner region and suppresses the density more efficiently at large radii, while preserving the finite total mass $\mathcal{M}_{\rm d}$. These properties make the Kuzmin-Toomre family a convenient set of analytic thin-disk solutions and a natural starting point for constructing the geometry of a compact object immersed in an accretion disk.

\subsection{Inverted Kuzmin-Toomre disks}

 While the Kuzmin-Toomre disks constitute a useful class of exact thin-disk solutions, their surface density remains non-vanishing at the origin. Since our ultimate goal is to model matter distributions surrounding a central BH, it is advantageous to work with annular disks possessing a central density depletion. This can be achieved by applying a Kelvin inversion~\cite{1993MNRAS.265..126B} to the Kuzmin-Toomre family.  The transformation generates a new class of finite-mass disks whose surface density vanishes at $\rho=0$, giving rise to annular configurations suitable for the superposition with a central BH. Explicitly under the Kelvin transformation implies, 
\begin{equation}
(\rho,z) \rightarrow \lt(\frac{b^2 \rho}{\rho^2+z^2},\; \frac{b^2 z}{\rho^2+z^2}\rt)~,
\end{equation}
under which the potential and density transform as
\begin{equation}
\nu \rightarrow \frac{b}{\sqrt{\rho^2+z^2}}\,\nu\;, \qquad
w(\rho) \rightarrow \frac{b^3}{\rho^3} w(b^2/\rho)~.
\end{equation}
The resulting potentials admit a closed-form expansion
\begin{align}\label{app:nu-i}
\nu_n^{(i)} =
-\binom{n+\frac{1}{2}}{n}
\frac{\mathcal{M}_{\rm d}}{(1+2n)!!}
\sum_{k=0}^{n}
\frac{(2n-k)!}{2^{n-k}(n-k)!}\nonumber \\
\times{_2F_1}(1+k,k-n;k-2n;2)
\frac{(-b)^k}{r_b^{k+1}} P_k(|\cos\theta_b|)~,
\end{align}
with surface density
\begin{align}\label{app:w-i}
w_n^{(i)}(\rho) =
\binom{n+\tfrac12}{n}
\frac{\mathcal{M}_{\rm d} b}{2\pi}
\frac{\rho^{2n}}{(\rho^2+b^2)^{n+3/2}}~.
\end{align}
These disks possess an annular structure, $w_n^{(i)} \propto \rho^{2n}$, ensuring vanishing density at the origin. The total mass is preserved under inversion.

The second metric function $\lambda$ can be written as a finite double series~\cite{Kotlarik:2022spo},
with coefficients involving hypergeometric functions,
\begin{align}\label{eq:lambda_inverted}
\lambda_n^{(i)} &=
-\binom{n+\tfrac{1}{2}}{n}^2
\frac{\mathcal{M}_{\rm d}^2 \sin^2\theta_b}{\big[(1+2n)!!\big]^2}\nonumber\\&\quad\times
\sum_{k,l=0}^{n}
B_{k,l}\;
\frac{(-b)^{k+l}}{r_b^{k+l+2}} \, P_{k,l}(\theta_b)~,
\end{align}
\begin{align}
B_{k,l} &=
\frac{(2n-k)!(2n-l)!}{2^{\,2n-k-l}(n-k)!(n-l)!(k+l+2)}
~,\nonumber\\
&\quad\times{}_2F_1(1+k,k-n;k-2n;2)\,\nonumber\\
&\quad\times
{}_2F_1(1+l,l-n;l-2n;2)~, \label{eq:Bkl}
\\\nonumber
P_{k,l} &=
(k+1)(l+1)P_k P_l
+ 2(k+1)\nonumber\\
&\quad\times|\cos\theta_b|\, P_k P'_l - \sin^2\theta_b\, P'_k P'_l, \label{eq:Pkl}
 \\ 
P'_k & \equiv
\frac{d}{d|\cos\theta_b|} P_k(|\cos\theta_b|)~. \label{eq:Pkprime}
\end{align}
The superscript $(i)$ in \cref{app:nu-i,app:w-i,eq:lambda_inverted} are indicative of the Kelvin inversion.\\

The inverted Kuzmin-Toomre family provides a convenient class of annular disks with finite total mass and vanishing central density. However, the parameter $n$ only controls the behaviour of the density near the origin through the factor $\rho^{2n}$. In contrast, the asymptotic falloff is universal,
\begin{equation}
w_n^{(i)} \sim \rho^{-3}~,
\qquad
\rho\rightarrow\infty~,
\end{equation}
for all members of the family. In applications where additional freedom in the radial mass distribution is desired, it is useful to consider a broader class of annular disks whose inner and outer profiles can be tuned independently. Such a generalization was constructed by Vogt and Letelier \cite{Vogt:2009sy} through suitable superpositions of Kuzmin-Toomre potentials.
\subsection{Vogt-Letelier disks}

 A more general class of thin-disk solutions can be constructed by superposing Kuzmin-Toomre potentials~\cite{Vogt:2009sy},
\begin{equation}
\nu^{(m,n)} = W_{(m,n)}
\sum_{k=0}^{n} \frac{(-1)^k}{\mathcal{M}_{\rm d}} \binom{n}{k}
\frac{\nu_{m+k}}{2m+2k+1}~,
\end{equation}
where the normalization constant
\begin{equation}
W_{(m,n)} = (2m+1)\binom{m+n+\tfrac12}{n}\mathcal{M}_{\rm d}
\end{equation}
ensures that the total mass remains $\mathcal{M}_{\rm d}$.

The corresponding surface density reads
\begin{equation}
w^{(m,n)}(\rho) =
W_{(m,n)}
\frac{ b^{2m+1}}{2\pi}
\frac{\rho^{2n}}{(\rho^2+b^2)^{m+n+3/2}}~,
\end{equation}
the parameters $m$ and $n$ control different aspects of the density profile. The factor $\rho^{2n}$ determines the degree of central depletion and therefore the annular character of the disk, while $m$ governs the asymptotic behaviour,
\begin{equation}
w^{(m,n)} \sim \rho^{-2m-3}~.
\end{equation}
Consequently, unlike the inverted Kuzmin-Toomre family, the Vogt-Letelier disks allow the inner and outer density profiles to be adjusted independently. The case $m=0$ reduces exactly to the inverted Kuzmin-Toomre disks.
The potentials can be expressed in closed form as
\begin{align}
\nu^{(m,n)} =
-(2m+1)\mathcal{M}_{\rm d}\binom{m+n+\tfrac12}{n} \nonumber\\
\times~ \sum_{j=0}^{m+n}Q_j^{(m,n)} \frac{b^j}{r_b^{j+1}} P_j(\cos\theta_b)~,
\end{align}
where the coefficients $Q_j^{(m,n)}$ are given in terms of generalized hypergeometric functions,
\begin{widetext}
\begin{align}
Q_j^{(m,n)} =
\begin{cases}
\displaystyle
\frac{2^{\,j-m}(2m-j)!}{(2m+1)!!(m-j)!}\,
{}_3F_2\!\lt(
\frac{2m-j+1}{2}, \frac{2m-j+2}{2}, -n;
\frac{2m+3}{2}, m-j+1; 1
\rt)\quad\quad\rm{if}\; j \le m~,
\\[10pt]
\displaystyle
\frac{(-1)^{j-m} j!}{(2j+1)!!}
\binom{n}{j-m}\,
{}_3F_2\!\lt(
\frac{j+1}{2}, \frac{j+2}{2}, j-m-n;
\frac{2j+3}{2}, j-m+1; 1
\rt)\quad\quad\rm{if}\;  j > m~.
\end{cases}
\end{align}
The corresponding metric function takes the form
\begin{align}
\lambda^{(m,n)} &=
-(2m+1)^2
\binom{m+n+\tfrac{1}{2}}{n}^2
\mathcal{M}_{\rm d}^2 \sin^2\theta_b\times
\sum_{k,l=0}^{m+n}
B^{(m,n)}_{k,l}
\frac{b^{k+l}}{r_b^{k+l+2}} \, P_{k,l}(\theta_b)~,
\label{eq:lambda_mn}
\\
B^{(m,n)}_{k,l} &=
\frac{Q^{(m,n)}_k \, Q^{(m,n)}_l}{k+l+2}~.
\label{eq:Bkl_mn}
\end{align}
The above families provide exact, fully relativistic thin-disk solutions with finite mass, smooth density profiles, and analytic expressions for both metric functions. The inverted Kuzmin-Toomre disks exhibit a universal outer decay and annular structure, while the Vogt-Letelier generalization allows additional control over the radial profile. These properties make them well-suited for constructing composite configurations involving a central BH~\cite{Kotlarik:2022spo}.

\section{Expression for $\Psi^{(1)}(r)$ for an accreting Schwarzschild BH}\label{app:psi-bh-1}

The computation of static LNs for accreting Schwarzschild BHs has been presented in \cref{LoveBHacc} in the main text. For this purpose, one needs to compute the solution to the master variable at linear-in-$\epsilon$ order. The corresponding solution reads,
\begin{align}
\Psi^{(1)}(r)=&\frac{1}{\lt(r_{\rm h} - 4 \mathbb{b}^2 r_{\rm h}\rt)^2} \bigg[2 r^2 \{14 \mathcal{A} + 3 (1 - 4 \mathbb{b}^2)^2 \mathbb{D}\} + 2 r \bigg(-2 \{10 - 45 \mathbb{b} + 660 \mathbb{b}^3 - 3456 \mathbb{b}^5 + 5760 \mathbb{b}^7\} r_{\rm h} \mathcal{A} \nonumber \\
&+ 32 (-1 + 6 \mathbb{b}^2) (1 - 21 \mathbb{b}^2 + 60 \mathbb{b}^4) \sqrt{r^2 - r r_{\rm h} + \mathbb{b}^2 r_{\rm h}^2} \mathcal{A} - 3 (1 - 4 \mathbb{b}^2)^2 r_{\rm h} \mathbb{D}\bigg) \nonumber \\ &+ r_{\rm h} \bigg(6 [2 - 15 \mathbb{b} + 4 \mathbb{b}^3 \{55 + 96 \mathbb{b}^2 (-3 + 5 \mathbb{b}^2)\}] r_{\rm h} \mathcal{A} - 32 (-1 + 6 \mathbb{b}^2) (1 - 21 \mathbb{b}^2 + 60 \mathbb{b}^4) \sqrt{r^2 - r r_{\rm h} + \mathbb{b}^2 r_{\rm h}^2} \mathcal{A} \nonumber \\  &+ (1 - 4 \mathbb{b}^2)^2 r_{\rm h} \mathbb{D}\bigg) +  2 (6 r^2 - 6 r r_{\rm h} + r_{\rm h}^2) \mathcal{A} \bigg\{\mathbb{b} \bigg[-15 + 4 \mathbb{b}^2 \{55 + 96 \mathbb{b}^2 (-3 + 5 \mathbb{b}^2)\}] [\log(-1 + 4 \mathbb{b}^2) \nonumber \\
&+ 2 \log(r/\{2 \mathbb{b} r - \mathbb{b} r_{\rm h} + \sqrt{r^2 - r r_{\rm h} + \mathbb{b}^2 r_{\rm h}^2}\})\bigg] + 2 \log\bigg(r/\{2 r - r_{\rm h} + 2 \sqrt{r^2 - r r_{\rm h} + \mathbb{b}^2 r_{\rm h}^2}\}\bigg)\bigg\}\bigg]~,
\end{align}
where $\mathcal{A}$ and $\mathbb{D}$ are arbitrary constants, and $r_{\rm h}=2M$ is the horizon of the isolated Schwarzschild BH.
\end{widetext}
\section{Green's function for accreting exotic compact objects}\label{App:Green}

In this appendix, we will discuss how to determine the Green's function for an accreting ECO, which boils down to solving \cref{eq:psi1}, and involves a non-trivial source term, originating from the vacuum perturbation equation, see \cref{eq:psi0}. The solution of \cref{eq:psi1} has been presented in \cref{eq:psi1G01} of the main text, which involves the Green's function, and is the one we will be focussing in this appendix. 

The computation of the Green's function requires imposing, either (a) boundary conditions at the boundaries, e.g., in the case of quasi-normal modes of a BH, one imposes a purely ingoing condition at the horizon and a purely outgoing condition at infinity, or, (b) imposing initial conditions. For LNs, there are no appropriate boundary conditions to be imposed at infinity, and hence we impose initial conditions at the surface of an ECO (for a BH, this will translate to regularity at the horizon). As our domain of interest is $r\in [r_{0},\infty)$, for $r_{0}\leq r<r'$, the Green's function $G_{<}(r,r')$ boils down to the homogeneous solution $\mathbb{A}P_{\ell}(r/M-1)+\mathbb{B}Q_{\ell}(r/M-1)$. For an ECO, we have a single initial condition: $G_{<}(r_{0},r')=0$ (we assume Dirichlet boundary condition throughout the main text and this appendix). However, the determination of the Green's function requires another initial condition, which we choose to be, $\partial_{r}G_{<}(r,r')|_{r=r_{0}}=\mathbb{C}$, where $\mathbb{C}$ is an arbitrary constant. As we will show, the choice of this constant will have no impact on the computation of the LNs. From \cref{eq:psi1G01}, upon imposing these initial conditions, we obtain, 
\begin{align}
&\mathbb{A}P_{\ell}(r_{0}/M-1)+\mathbb{B}Q_{\ell}(r_{0}/M-1)=0\,,
\\
&\mathbb{A}P'_{\ell}(r_{0}/M-1)+\mathbb{B}Q_{\ell}'(r_{0}/M-1)=\mathbb{C}\,,
\end{align}
where, `prime' denotes the derivative with respect to the radial coordinate $r$. Thus, we obtain, 
\begin{align}
&\frac{\mathbb{B}}{\mathbb{A}}=-\frac{P_{\ell}(r_{0}/M-1)}{Q_{\ell}(r_{0}/M-1)}\,,
\\
&\mathbb{A}=\frac{\mathbb{C}Q_{\ell}(r_{0}/M-1)}{W_{0}}\,,
\end{align}
where, $W_{0}$ is the Wronskian of the homogeneous solution, evaluated at the boundary $r=r_{0}$. which turns out to be, 
\begin{align}
W_{0}&\equiv Q_{\ell}(r_{0}/M-1)P'_{\ell}(r_{0}/M-1)
\nonumber
\\
&\quad-P_{\ell}(r_{0}/M-1)Q_{\ell}'(r_{0}/M-1)=\frac{M}{r_{0}^{2}f(r_{0})}\,.
\end{align}
Note that the Wronskian is non-zero throughout the region of interest, $r\in[r_{0},\infty)$, as expected. Thus for $r_{0}\leq r<r'$, the Green's function becomes, 
\begin{align}
G_{<}(r,r')&=\frac{\mathbb{C}}{W_{0}}\Big[Q_{\ell}(r_{0}/M-1)P_{\ell}(r/M-1)
\nonumber
\\
&\quad-P_{\ell}(r_{0}/M-1)Q_{\ell}(r/M-1)\Big]\,.
\end{align}
For, $r>r'$, on the other hand, we use the continuity of the Green's function at $r=r'$ and the discontinuity in its first derivative at $r=r'$, to arrive at the following conditions:
\begin{align}
&\mathcal{A}P_{\ell}(r'/M-1)+\mathcal{B}Q_{\ell}(r'/M-1)
\nonumber
\\
&\quad=\frac{\mathbb{C}}{W_{0}}\Big[Q_{\ell}(r_{0}/M-1)P_{\ell}(r'/M-1)
\nonumber
\\
&\qquad-P_{\ell}(r_{0}/M-1)Q_{\ell}(r'/M-1)\Big]
\label{solgrt01}
\\
&\mathcal{A}P'_{\ell}(r'/M-1)+\mathcal{B}Q'_{\ell}(r'/M-1)
\nonumber
\\
&\quad-\frac{\mathbb{C}}{W_{0}}\Big[Q_{\ell}(r_{0}/M-1)P'_{\ell}(r'/M-1)
\nonumber
\\
&\qquad-P_{\ell}(r_{0}/M-1)Q'_{\ell}(r'/M-1)\Big]=\frac{1}{r'^{2}f(r')}\,,
\label{solgrt02}
\end{align}
where, we have expressed the Green's function for $r>r'$, as, $G_{>}(r,r')=\mathcal{A}P_{\ell}(r/M-1)+\mathcal{B}Q_{\ell}(r/M-1)$. The arbitary constants $\mathcal{A}$ and $\mathcal{B}$ can be determined by solving \cref{solgrt01} and \cref{solgrt02}, which yields,
\begin{align}
\mathcal{A}&=\frac{\mathbb{C}}{W_0}Q_{\ell}(r_0/M-1)+\frac{Q_{\ell}(r'/M-1)}{M}~,
\\
\mathcal{B}&=-\frac{\mathbb{C}}{W_0}P_{\ell}(r_0/M-1)-\frac{P_{\ell}(r'/M-1)}{M}~.
\end{align}
Thus, using these two results, we can obtain the Green's function $G_{>}(r,r')$. Combining both $G_{<}$ and $G_{>}$, we finally obtain the complete Green's function for $r\in [r_{0},\infty)$ as,
\begin{widetext}
\begin{align}
G(r,r')=&\frac{\mathbb{C}}{W_0}\Big[Q_{\ell}(r_{0}/M-1)P_{\ell}(r/M-1)-P_{\ell}(r_{0}/M-1)Q_{\ell}(r/M-1)\Big]
\nonumber
\\
&+\Theta(r-r')\frac{Q_{\ell}(r'/M-1)P_{\ell}(r/M-1)-P_{\ell}(r'/M-1)Q_{\ell}(r/M-1)}{M}\,,
\end{align}
\end{widetext}
where, $\Theta(r-r')=1$ for $r>r'$ and vanishes elsewhere. 

Using the above form of the Green's function over the entire domain of interest, the inhomogeneous part of the solution, as presented in \cref{eq:psi1G01} becomes,
\begin{align}
&\Psi^{(1)}_{\ell m_z}|_{\rm inhomogeneous}=\int_{r_0}^{\infty}G(r,r')\,\mathbb{V}(r')\Psi^{(0)}_{\ell m_{z}}(r')\,dr' 
\nonumber
\\
&=\frac{P_\ell(r/M-1)}{M}\int_{r_0}^{r}Q_\ell(r'/M-1)\,\mathbb{V}(r')\Psi^{(0)}_{\ell m_{z}}(r')\,dr'\nonumber
\\
&-\frac{Q_\ell(r/M-1)}{M}\int_{r_0}^{r}P_\ell(r'/M-1)\,\mathbb{V}(r')\Psi^{(0)}_{\ell m_{z}}(r')\,dr'
\nonumber
\\
&+{\frac{\lambda_{0}}{W_0 \mathbb{A}}}\Psi^{(0)}_{\ell m_z}(r)\,,
\end{align}
where, we have introduced the constant $\lambda_{0}$, dependent on the location of the surface of the ECO $r_{0}$, which has the following expression,
\begin{align}
\lambda_{0}\equiv \int_{r_0}^{\infty}\mathbb{V}(r')\Psi^{(0)}_{\ell m_z}(r')\,dr'\,.
\end{align}
These are the results that we have used in the main text. Note that, the final result does not depend on the choice of the arbitrary constant $\mathbb{C}$. 

\bibliography{ref}

\providecommand{\href}[2]{#2}\begingroup\raggedright\begin{thebibliography}{10}

\bibitem{1909MNRAS..69..476L}
A.~E.~H. {Love}, ``{The yielding of the Earth to disturbing forces},'' \href{http://dx.doi.org/10.1093/mnras/69.6.476}{{\em \mnras} {\bfseries 69} (Apr., 1909) 476}.

\bibitem{10.1098/rspa.1909.0008}
A.~E.~H. Love, ``The yielding of the earth to disturbing forces,'' \href{http://dx.doi.org/10.1098/rspa.1909.0008}{{\em Proceedings of the Royal Society of London. Series A, Containing Papers of a Mathematical and Physical Character} {\bfseries 82} no.~551, (02, 1909) 73--88}.

\bibitem{1911spge.book.....L}
A.~E.~H. {Love}, ``{Some Problems of Geodynamics},'' 1911.

\bibitem{Hinderer:2007mb}
T.~Hinderer, ``{Tidal Love numbers of neutron stars},'' \href{http://dx.doi.org/10.1086/533487}{{\em Astrophys. J.} {\bfseries 677} (2008) 1216--1220}, \href{http://arxiv.org/abs/0711.2420}{{\ttfamily arXiv:0711.2420 [astro-ph]}}. [Erratum: Astrophys.J. 697, 964 (2009)].

\bibitem{Binnington:2009bb}
T.~Binnington and E.~Poisson, ``{Relativistic theory of tidal Love numbers},'' \href{http://dx.doi.org/10.1103/PhysRevD.80.084018}{{\em Phys. Rev. D} {\bfseries 80} (2009) 084018}, \href{http://arxiv.org/abs/0906.1366}{{\ttfamily arXiv:0906.1366 [gr-qc]}}.

\bibitem{Damour:2009vw}
T.~Damour and A.~Nagar, ``{Relativistic tidal properties of neutron stars},'' \href{http://dx.doi.org/10.1103/PhysRevD.80.084035}{{\em Phys. Rev. D} {\bfseries 80} (2009) 084035}, \href{http://arxiv.org/abs/0906.0096}{{\ttfamily arXiv:0906.0096 [gr-qc]}}.

\bibitem{Flanagan:2007ix}
E.~E. Flanagan and T.~Hinderer, ``{Constraining neutron star tidal Love numbers with gravitational wave detectors},'' \href{http://dx.doi.org/10.1103/PhysRevD.77.021502}{{\em Phys. Rev. D} {\bfseries 77} (2008) 021502}, \href{http://arxiv.org/abs/0709.1915}{{\ttfamily arXiv:0709.1915 [astro-ph]}}.

\bibitem{Yagi:2016bkt}
K.~Yagi and N.~Yunes, ``{Approximate Universal Relations for Neutron Stars and Quark Stars},'' \href{http://dx.doi.org/10.1016/j.physrep.2017.03.002}{{\em Phys. Rept.} {\bfseries 681} (2017) 1--72}, \href{http://arxiv.org/abs/1608.02582}{{\ttfamily arXiv:1608.02582 [gr-qc]}}.

\bibitem{Yagi:2013bca}
K.~Yagi and N.~Yunes, ``{I-Love-Q},'' \href{http://dx.doi.org/10.1126/science.1236462}{{\em Science} {\bfseries 341} (2013) 365--368}, \href{http://arxiv.org/abs/1302.4499}{{\ttfamily arXiv:1302.4499 [gr-qc]}}.

\bibitem{Yagi:2013awa}
K.~Yagi and N.~Yunes, ``{I-Love-Q Relations in Neutron Stars and their Applications to Astrophysics, Gravitational Waves and Fundamental Physics},'' \href{http://dx.doi.org/10.1103/PhysRevD.88.023009}{{\em Phys. Rev. D} {\bfseries 88} no.~2, (2013) 023009}, \href{http://arxiv.org/abs/1303.1528}{{\ttfamily arXiv:1303.1528 [gr-qc]}}.

\bibitem{Cardoso:2017cfl}
V.~Cardoso, E.~Franzin, A.~Maselli, P.~Pani, and G.~Raposo, ``{Testing strong-field gravity with tidal Love numbers},'' \href{http://dx.doi.org/10.1103/PhysRevD.95.084014}{{\em Phys. Rev. D} {\bfseries 95} no.~8, (2017) 084014}, \href{http://arxiv.org/abs/1701.01116}{{\ttfamily arXiv:1701.01116 [gr-qc]}}. [Addendum: Phys.Rev.D 95, 089901 (2017)].

\bibitem{Abbott:2018exr}
{\bfseries LIGO Scientific, Virgo} Collaboration, B.~P. Abbott {\em et~al.}, ``{GW170817: Measurements of neutron star radii and equation of state},'' \href{http://dx.doi.org/10.1103/PhysRevLett.121.161101}{{\em Phys. Rev. Lett.} {\bfseries 121} no.~16, (2018) 161101},
\href{http://arxiv.org/abs/1805.11581}{{\ttfamily arXiv:1805.11581 [gr-qc]}}.

\bibitem{LIGOScientific:2017vwq}
{\bfseries LIGO Scientific, Virgo} Collaboration, B.~P. Abbott {\em et~al.}, ``{GW170817: Observation of Gravitational Waves from a Binary Neutron Star Inspiral},'' \href{http://dx.doi.org/10.1103/PhysRevLett.119.161101}{{\em Phys. Rev. Lett.} {\bfseries 119} no.~16, (2017) 161101}, \href{http://arxiv.org/abs/1710.05832}{{\ttfamily arXiv:1710.05832 [gr-qc]}}.

\bibitem{Abbott:2018wiz}
{\bfseries LIGO Scientific, Virgo} Collaboration, B.~P. Abbott {\em et~al.}, ``{Properties of the binary neutron star merger GW170817},'' \href{http://dx.doi.org/10.1103/PhysRevX.9.011001}{{\em Phys. Rev.} {\bfseries X9} no.~1, (2019) 011001},
\href{http://arxiv.org/abs/1805.11579}{{\ttfamily arXiv:1805.11579 [gr-qc]}}.

\bibitem{Chakraborty:2026qru}
S.~Chakraborty and P.~Pani, ``{Tidal Response of Compact Objects},'' \href{http://arxiv.org/abs/2604.08679}{{\ttfamily arXiv:2604.08679 [gr-qc]}}.

\bibitem{Rodriguez:2026iot}
M.~J. Rodr{\'\i}guez, L.~Santoni, and A.~R. Solomon, ``{Love numbers of black holes and compact objects},'' \href{http://arxiv.org/abs/2604.08653}{{\ttfamily arXiv:2604.08653 [gr-qc]}}.

\bibitem{Kol:2011vg}
B.~Kol and M.~Smolkin, ``{Black hole stereotyping: Induced gravito-static polarization},'' \href{http://dx.doi.org/10.1007/JHEP02(2012)010}{{\em JHEP} {\bfseries 02} (2012) 010}, \href{http://arxiv.org/abs/1110.3764}{{\ttfamily arXiv:1110.3764 [hep-th]}}.

\bibitem{2013arXiv1304.2228C}
S.~Chakrabarti, T.~Delsate, and J.~Steinhoff, ``{New perspectives on neutron star and black hole spectroscopy and dynamic tides},'' \href{http://arxiv.org/abs/1304.2228}{{\ttfamily arXiv:1304.2228 [gr-qc]}}.

\bibitem{Gurlebeck:2015xpa}
N.~G\"urlebeck, ``{No-hair theorem for Black Holes in Astrophysical Environments},'' \href{http://dx.doi.org/10.1103/PhysRevLett.114.151102}{{\em Phys. Rev. Lett.} {\bfseries 114} no.~15, (2015) 151102}, \href{http://arxiv.org/abs/1503.03240}{{\ttfamily arXiv:1503.03240 [gr-qc]}}.

\bibitem{Chia:2020yla}
H.~S. Chia, ``{Tidal deformation and dissipation of rotating black holes},'' \href{http://dx.doi.org/10.1103/PhysRevD.104.024013}{{\em Phys. Rev. D} {\bfseries 104} no.~2, (2021) 024013}, \href{http://arxiv.org/abs/2010.07300}{{\ttfamily arXiv:2010.07300 [gr-qc]}}.

\bibitem{Bhatt:2023zsy}
R.~P. Bhatt, S.~Chakraborty, and S.~Bose, ``{Addressing issues in defining the Love numbers for black holes},'' \href{http://dx.doi.org/10.1103/PhysRevD.108.084013}{{\em Phys. Rev. D} {\bfseries 108} no.~8, (2023) 084013}, \href{http://arxiv.org/abs/2306.13627}{{\ttfamily arXiv:2306.13627 [gr-qc]}}.

\bibitem{LeTiec:2020bos}
A.~Le~Tiec, M.~Casals, and E.~Franzin, ``{Tidal Love Numbers of Kerr Black Holes},'' \href{http://dx.doi.org/10.1103/PhysRevD.103.084021}{{\em Phys. Rev. D} {\bfseries 103} no.~8, (2021) 084021}, \href{http://arxiv.org/abs/2010.15795}{{\ttfamily arXiv:2010.15795 [gr-qc]}}.

\bibitem{Charalambous:2021mea}
P.~Charalambous, S.~Dubovsky, and M.~M. Ivanov, ``{On the Vanishing of Love Numbers for Kerr Black Holes},'' \href{http://dx.doi.org/10.1007/JHEP05(2021)038}{{\em JHEP} {\bfseries 05} (2021) 038}, \href{http://arxiv.org/abs/2102.08917}{{\ttfamily arXiv:2102.08917 [hep-th]}}.

\bibitem{Chakraborty:2025zyb}
S.~Chakraborty, P.~Heidmann, and P.~Pani, ``{Fermionic response of black holes in general relativity},'' \href{http://dx.doi.org/10.1103/2yr1-9ymw}{{\em Phys. Rev. D} {\bfseries 113} no.~6, (2026) L061503}, \href{http://arxiv.org/abs/2508.20155}{{\ttfamily arXiv:2508.20155 [gr-qc]}}.

\bibitem{hui2022ladder-678}
L.~Hui, A.~Joyce, R.~Penco, L.~Santoni, and A.~R. Solomon, ``{Ladder symmetries of black holes. Implications for love numbers and no-hair theorems},'' \href{http://dx.doi.org/10.1088/1475-7516/2022/01/032}{{\em JCAP} {\bfseries 01} no.~01, (2022) 032}, \href{http://arxiv.org/abs/2105.01069}{{\ttfamily arXiv:2105.01069 [hep-th]}}.

\bibitem{achour2022hidden-8c7}
J.~Ben~Achour, E.~R. Livine, S.~Mukohyama, and J.-P. Uzan, ``{Hidden symmetry of the static response of black holes: applications to Love numbers},'' \href{http://dx.doi.org/10.1007/JHEP07(2022)112}{{\em JHEP} {\bfseries 07} (2022) 112}, \href{http://arxiv.org/abs/2202.12828}{{\ttfamily arXiv:2202.12828 [gr-qc]}}.

\bibitem{charalambous2021hidden-5e0}
P.~Charalambous, S.~Dubovsky, and M.~M. Ivanov, ``{Hidden Symmetry of Vanishing Love Numbers},'' \href{http://dx.doi.org/10.1103/PhysRevLett.127.101101}{{\em Phys. Rev. Lett.} {\bfseries 127} no.~10, (2021) 101101}, \href{http://arxiv.org/abs/2103.01234}{{\ttfamily arXiv:2103.01234 [hep-th]}}.

\bibitem{ivanov2023vanishing-9aa}
M.~M. Ivanov and Z.~Zhou, ``{Vanishing of Black Hole Tidal Love Numbers from Scattering Amplitudes},'' \href{http://dx.doi.org/10.1103/PhysRevLett.130.091403}{{\em Phys. Rev. Lett.} {\bfseries 130} no.~9, (2023) 091403}, \href{http://arxiv.org/abs/2209.14324}{{\ttfamily arXiv:2209.14324 [hep-th]}}.

\bibitem{creci2021tidal-42e}
G.~Creci, T.~Hinderer, and J.~Steinhoff, ``{Tidal response from scattering and the role of analytic continuation},'' \href{http://dx.doi.org/10.1103/PhysRevD.104.124061}{{\em Phys. Rev. D} {\bfseries 104} no.~12, (2021) 124061}, \href{http://arxiv.org/abs/2108.03385}{{\ttfamily arXiv:2108.03385 [gr-qc]}}. [Erratum: Phys.Rev.D 105, 109902 (2022)].

\bibitem{Cardoso:2019vof}
V.~Cardoso, L.~Gualtieri, and C.~J. Moore, ``{Gravitational waves and higher dimensions: Love numbers and Kaluza-Klein excitations},'' \href{http://dx.doi.org/10.1103/PhysRevD.100.124037}{{\em Phys. Rev. D} {\bfseries 100} no.~12, (2019) 124037}, \href{http://arxiv.org/abs/1910.09557}{{\ttfamily arXiv:1910.09557 [gr-qc]}}.

\bibitem{Hui:2020xxx}
L.~Hui, A.~Joyce, R.~Penco, L.~Santoni, and A.~R. Solomon, ``{Static response and Love numbers of Schwarzschild black holes},'' \href{http://dx.doi.org/10.1088/1475-7516/2021/04/052}{{\em JCAP} {\bfseries 04} (2021) 052}, \href{http://arxiv.org/abs/2010.00593}{{\ttfamily arXiv:2010.00593 [hep-th]}}.

\bibitem{DeLuca:2024ufn}
V.~De~Luca, A.~Garoffolo, J.~Khoury, and M.~Trodden, ``{Tidal Love numbers and Green\textquoteright{}s functions in black hole spacetimes},'' \href{http://dx.doi.org/10.1103/PhysRevD.110.064081}{{\em Phys. Rev. D} {\bfseries 110} no.~6, (2024) 064081}, \href{http://arxiv.org/abs/2407.07156}{{\ttfamily arXiv:2407.07156 [gr-qc]}}.

\bibitem{Bhatt:2024mvr}
R.~P. Bhatt and C.~Singha, ``{Scalar tidal response of a rotating BTZ black hole},'' \href{http://dx.doi.org/10.1007/JHEP11(2024)154}{{\em JHEP} {\bfseries 11} (2024) 154}, \href{http://arxiv.org/abs/2407.09470}{{\ttfamily arXiv:2407.09470 [gr-qc]}}.

\bibitem{Chakravarti:2018vlt}
K.~Chakravarti, S.~Chakraborty, S.~Bose, and S.~SenGupta, ``{Tidal Love numbers of black holes and neutron stars in the presence of higher dimensions: Implications of GW170817},'' \href{http://dx.doi.org/10.1103/PhysRevD.99.024036}{{\em Phys. Rev. D} {\bfseries 99} no.~2, (2019) 024036}, \href{http://arxiv.org/abs/1811.11364}{{\ttfamily arXiv:1811.11364 [gr-qc]}}.

\bibitem{Rodriguez:2023xjd}
M.~J. Rodriguez, L.~Santoni, A.~R. Solomon, and L.~F. Temoche, ``{Love numbers for rotating black holes in higher dimensions},'' \href{http://dx.doi.org/10.1103/PhysRevD.108.084011}{{\em Phys. Rev. D} {\bfseries 108} no.~8, (2023) 084011}, \href{http://arxiv.org/abs/2304.03743}{{\ttfamily arXiv:2304.03743 [hep-th]}}.

\bibitem{Cardoso:2018ptl}
V.~Cardoso, M.~Kimura, A.~Maselli, and L.~Senatore, ``{Black Holes in an Effective Field Theory Extension of General Relativity},'' \href{http://dx.doi.org/10.1103/PhysRevLett.121.251105}{{\em Phys. Rev. Lett.} {\bfseries 121} no.~25, (2018) 251105}, \href{http://arxiv.org/abs/1808.08962}{{\ttfamily arXiv:1808.08962 [gr-qc]}}. [Erratum: Phys.Rev.Lett. 131, 109903 (2023)].

\bibitem{DeLuca:2022tkm}
V.~De~Luca, J.~Khoury, and S.~S.~C. Wong, ``{Implications of the weak gravity conjecture for tidal Love numbers of black holes},'' \href{http://dx.doi.org/10.1103/PhysRevD.108.044066}{{\em Phys. Rev. D} {\bfseries 108} no.~4, (2023) 044066}, \href{http://arxiv.org/abs/2211.14325}{{\ttfamily arXiv:2211.14325 [hep-th]}}.

\bibitem{Singha:2025xah}
C.~Singha and S.~Chakraborty, ``{Tidal deformation of black holes in Lovelock gravity},'' \href{http://dx.doi.org/10.1103/8jfh-9rb6}{{\em Phys. Rev. D} {\bfseries 113} no.~2, (2026) 024005}, \href{http://arxiv.org/abs/2508.14944}{{\ttfamily arXiv:2508.14944 [gr-qc]}}.

\bibitem{Barbosa:2025uau}
S.~Barbosa, P.~Brax, S.~Fichet, and L.~de~Souza, ``{Running Love numbers and the Effective Field Theory of gravity},'' \href{http://dx.doi.org/10.1088/1475-7516/2025/07/071}{{\em JCAP} {\bfseries 07} (2025) 071}, \href{http://arxiv.org/abs/2501.18684}{{\ttfamily arXiv:2501.18684 [hep-th]}}.

\bibitem{Barbosa:2026qcv}
S.~Barbosa, S.~Fichet, and L.~de~Souza, ``{Running Love numbers of charged black holes},'' \href{http://dx.doi.org/10.1088/1475-7516/2026/09/009}{{\em JCAP} {\bfseries 09} (2026) 009}, \href{http://arxiv.org/abs/2602.00349}{{\ttfamily arXiv:2602.00349 [hep-th]}}.

\bibitem{Emparan:2017qxd}
R.~Emparan, A.~Fernandez-Pique, and R.~Luna, ``{Geometric polarization of plasmas and Love numbers of AdS black branes},'' \href{http://dx.doi.org/10.1007/JHEP09(2017)150}{{\em JHEP} {\bfseries 09} (2017) 150}, \href{http://arxiv.org/abs/1707.02777}{{\ttfamily arXiv:1707.02777 [hep-th]}}.

\bibitem{Nair:2024mya}
S.~Nair, S.~Chakraborty, and S.~Sarkar, ``{Asymptotically de Sitter black holes have nonzero tidal Love numbers},'' \href{http://dx.doi.org/10.1103/PhysRevD.109.064025}{{\em Phys. Rev. D} {\bfseries 109} no.~6, (2024) 064025}, \href{http://arxiv.org/abs/2401.06467}{{\ttfamily arXiv:2401.06467 [gr-qc]}}.

\bibitem{Franzin:2024cah}
E.~Franzin, A.~M. Frassino, and J.~V. Rocha, ``{Tidal Love numbers of static black holes in anti-de Sitter},'' \href{http://dx.doi.org/10.1007/JHEP12(2024)224}{{\em JHEP} {\bfseries 12} (2025) 224}, \href{http://arxiv.org/abs/2410.23545}{{\ttfamily arXiv:2410.23545 [hep-th]}}.

\bibitem{Silvestrini:2025lbe}
M.~Silvestrini, E.~Maggio, S.~Chakraborty, and P.~Pani, ``{Tidal deformations of compact objects from the membrane paradigm},'' \href{http://dx.doi.org/10.1103/pky4-23jb}{{\em Phys. Rev. D} {\bfseries 112} no.~12, (2025) 124021}, \href{http://arxiv.org/abs/2506.16516}{{\ttfamily arXiv:2506.16516 [gr-qc]}}.

\bibitem{Chakraborty:2023zed}
S.~Chakraborty, E.~Maggio, M.~Silvestrini, and P.~Pani, ``{Dynamical tidal Love numbers of Kerr-like compact objects},'' \href{http://dx.doi.org/10.1103/PhysRevD.110.084042}{{\em Phys. Rev. D} {\bfseries 110} no.~8, (2024) 084042}, \href{http://arxiv.org/abs/2310.06023}{{\ttfamily arXiv:2310.06023 [gr-qc]}}.

\bibitem{Nair:2022xfm}
S.~Nair, S.~Chakraborty, and S.~Sarkar, ``{Dynamical Love numbers for area quantized black holes},'' \href{http://dx.doi.org/10.1103/PhysRevD.107.124041}{{\em Phys. Rev. D} {\bfseries 107} no.~12, (2023) 124041}, \href{http://arxiv.org/abs/2208.06235}{{\ttfamily arXiv:2208.06235 [gr-qc]}}.

\bibitem{Pani:2015hfa}
P.~Pani, L.~Gualtieri, A.~Maselli, and V.~Ferrari, ``{Tidal deformations of a spinning compact object},'' \href{http://dx.doi.org/10.1103/PhysRevD.92.024010}{{\em Phys. Rev. D} {\bfseries 92} no.~2, (2015) 024010}, \href{http://arxiv.org/abs/1503.07365}{{\ttfamily arXiv:1503.07365 [gr-qc]}}.

\bibitem{Chakraborty:2025wvs}
S.~Chakraborty, V.~De~Luca, L.~Gualtieri, and P.~Pani, ``{Dynamical Love numbers of black holes: Theory and gravitational waveforms},'' \href{http://dx.doi.org/10.1103/fr3y-s1sz}{{\em Phys. Rev. D} {\bfseries 112} no.~10, (2025) 104015}, \href{http://arxiv.org/abs/2507.22994}{{\ttfamily arXiv:2507.22994 [gr-qc]}}.

\bibitem{Bhatt:2024yyz}
R.~P. Bhatt, S.~Chakraborty, and S.~Bose, ``{Rotating black holes experience dynamical tides},'' \href{http://dx.doi.org/10.1103/PhysRevD.111.L041504}{{\em Phys. Rev. D} {\bfseries 111} no.~4, (2025) L041504}, \href{http://arxiv.org/abs/2406.09543}{{\ttfamily arXiv:2406.09543 [gr-qc]}}.

\bibitem{katagiri2024relativistic-5c0}
T.~Katagiri, K.~Yagi, and V.~Cardoso, ``{Relativistic dynamical tides: Subtleties and calibration},'' \href{http://dx.doi.org/10.1103/PhysRevD.111.084080}{{\em Phys. Rev. D} {\bfseries 111} no.~8, (2025) 084080}, \href{http://arxiv.org/abs/2409.18034}{{\ttfamily arXiv:2409.18034 [gr-qc]}}.

\bibitem{saketh2023dynamical-b30}
M.~V.~S. Saketh, Z.~Zhou, and M.~M. Ivanov, ``{Dynamical tidal response of Kerr black holes from scattering amplitudes},'' \href{http://dx.doi.org/10.1103/PhysRevD.109.064058}{{\em Phys. Rev. D} {\bfseries 109} no.~6, (2024) 064058}, \href{http://arxiv.org/abs/2307.10391}{{\ttfamily arXiv:2307.10391 [hep-th]}}.

\bibitem{Kehagias:2024rtz}
A.~Kehagias and A.~Riotto, ``{Black holes in a gravitational field: the non-linear static love number of Schwarzschild black holes vanishes},'' \href{http://dx.doi.org/10.1088/1475-7516/2025/05/039}{{\em JCAP} {\bfseries 05} (2025) 039}, \href{http://arxiv.org/abs/2410.11014}{{\ttfamily arXiv:2410.11014 [gr-qc]}}.

\bibitem{Chakraborty:2026dox}
S.~Chakraborty, M.~V.~S. Saketh, T.~Hinderer, and J.~Steinhoff, ``{Dynamical tidal Love numbers of black holes under generic perturbations: Connecting black hole perturbation theory with effective field theory},'' \href{http://arxiv.org/abs/2605.00693}{{\ttfamily arXiv:2605.00693 [gr-qc]}}.

\bibitem{Perry:2024vwz}
M.~Perry and M.~J. Rodriguez, ``{Love numbers for extremal Kerr black holes},'' \href{http://dx.doi.org/10.1103/s5vk-vdg1}{{\em Phys. Rev. D} {\bfseries 112} no.~12, (2025) 126004}, \href{http://arxiv.org/abs/2412.19699}{{\ttfamily arXiv:2412.19699 [hep-th]}}.

\bibitem{Ghosh:2026vig}
R.~Ghosh, R.~P. Bhatt, S.~Chakraborty, and S.~Bose, ``{Universal Ladder Structure Across Scales: From Quantum to Black Hole Physics},'' \href{http://arxiv.org/abs/2604.06249}{{\ttfamily arXiv:2604.06249 [gr-qc]}}.

\bibitem{Chakraborty:2024gcr}
S.~Chakraborty, G.~Comp{\`e}re, and L.~Machet, ``{Tidal Love numbers and quasinormal modes of the Schwarzschild-Hernquist black hole},'' \href{http://dx.doi.org/10.1103/4p2c-rwdh}{{\em Phys. Rev. D} {\bfseries 112} no.~2, (2025) 024015}, \href{http://arxiv.org/abs/2412.14831}{{\ttfamily arXiv:2412.14831 [gr-qc]}}.

\bibitem{Chakravarti:2025awj}
K.~Chakravarti and C.~Singha, ``{Tidal Love numbers and quasi-normal modes of the ECO in a Dark Matter halo},'' \href{http://arxiv.org/abs/2509.03556}{{\ttfamily arXiv:2509.03556 [gr-qc]}}.

\bibitem{Cardoso:2019upw}
V.~Cardoso and F.~Duque, ``{Environmental effects in gravitational-wave physics: Tidal deformability of black holes immersed in matter},'' \href{http://dx.doi.org/10.1103/PhysRevD.101.064028}{{\em Phys. Rev. D} {\bfseries 101} no.~6, (2020) 064028}, \href{http://arxiv.org/abs/1912.07616}{{\ttfamily arXiv:1912.07616 [gr-qc]}}.

\bibitem{Cardoso:2021wlq}
V.~Cardoso, K.~Destounis, F.~Duque, R.~P. Macedo, and A.~Maselli, ``{Black holes in galaxies: Environmental impact on gravitational-wave generation and propagation},'' \href{http://dx.doi.org/10.1103/PhysRevD.105.L061501}{{\em Phys. Rev. D} {\bfseries 105} no.~6, (2022) L061501}, \href{http://arxiv.org/abs/2109.00005}{{\ttfamily arXiv:2109.00005 [gr-qc]}}.

\bibitem{DOnofrio:2026ulh}
S.~D'Onofrio, S.~Datta, and A.~Maselli, ``{Axial tidal Love numbers of black holes in matter environments},'' \href{http://arxiv.org/abs/2605.02633}{{\ttfamily arXiv:2605.02633 [gr-qc]}}.

\bibitem{Zhao:2026eti}
Y.-Q. Zhao and P.~Pani, ``{Quasinormal modes and tidal responses of black holes in generic anisotropic matter environments},'' \href{http://arxiv.org/abs/2606.11380}{{\ttfamily arXiv:2606.11380 [gr-qc]}}.

\bibitem{Cannizzaro:2024fpz}
E.~Cannizzaro, V.~De~Luca, and P.~Pani, ``{Tidal deformability of black holes surrounded by thin accretion disks},'' \href{http://dx.doi.org/10.1103/PhysRevD.110.123004}{{\em Phys. Rev. D} {\bfseries 110} no.~12, (2024) 123004}, \href{http://arxiv.org/abs/2408.14208}{{\ttfamily arXiv:2408.14208 [astro-ph.HE]}}.

\bibitem{Cardoso:2016oxy}
V.~Cardoso, S.~Hopper, C.~F.~B. Macedo, C.~Palenzuela, and P.~Pani, ``{Gravitational-wave signatures of exotic compact objects and of quantum corrections at the horizon scale},'' \href{http://dx.doi.org/10.1103/PhysRevD.94.084031}{{\em Phys. Rev. D} {\bfseries 94} no.~8, (2016) 084031}, \href{http://arxiv.org/abs/1608.08637}{{\ttfamily arXiv:1608.08637 [gr-qc]}}.

\bibitem{Cardoso:2016rao}
V.~Cardoso, E.~Franzin, and P.~Pani, ``{Is the gravitational-wave ringdown a probe of the event horizon?},'' \href{http://dx.doi.org/10.1103/PhysRevLett.116.171101}{{\em Phys. Rev. Lett.} {\bfseries 116} no.~17, (2016) 171101}, \href{http://arxiv.org/abs/1602.07309}{{\ttfamily arXiv:1602.07309 [gr-qc]}}. [Erratum: Phys.Rev.Lett. 117, 089902 (2016)].

\bibitem{Bueno:2017hyj}
P.~Bueno, P.~A. Cano, F.~Goelen, T.~Hertog, and B.~Vercnocke, ``{Echoes of Kerr-like wormholes},'' \href{http://dx.doi.org/10.1103/PhysRevD.97.024040}{{\em Phys. Rev. D} {\bfseries 97} no.~2, (2018) 024040}, \href{http://arxiv.org/abs/1711.00391}{{\ttfamily arXiv:1711.00391 [gr-qc]}}.

\bibitem{Cardoso:2019rvt}
V.~Cardoso and P.~Pani, ``{Testing the nature of dark compact objects: a status report},'' \href{http://dx.doi.org/10.1007/s41114-019-0020-4}{{\em Living Rev. Relativ.} {\bfseries 22} no.~1, (2019) 4}, \href{http://arxiv.org/abs/1904.05363}{{\ttfamily arXiv:1904.05363 [gr-qc]}}.

\bibitem{Maggio:2021ans}
E.~Maggio, P.~Pani, and G.~Raposo, \href{http://dx.doi.org/10.1007/978-981-15-4702-7_29-1}{``{Testing the nature of dark compact objects with gravitational waves},''} in {\em Handbook of Gravitational Wave Astronomy}, C.~Bambi, S.~Katsanevas, and K.~D. Kokkotas, eds.
\newblock Springer, Singapore, 5, 2021.
\newblock \href{http://arxiv.org/abs/2105.06410}{{\ttfamily arXiv:2105.06410 [gr-qc]}}.

\bibitem{DelGrosso:2023trq}
L.~Del~Grosso, G.~Franciolini, P.~Pani, and A.~Urbano, ``{Fermion soliton stars},'' \href{http://dx.doi.org/10.1103/PhysRevD.108.044024}{{\em Phys. Rev. D} {\bfseries 108} no.~4, (2023) 044024}, \href{http://arxiv.org/abs/2301.08709}{{\ttfamily arXiv:2301.08709 [gr-qc]}}.

\bibitem{Pani:2010em}
P.~Pani, E.~Berti, V.~Cardoso, Y.~Chen, and R.~Norte, ``{Gravitational-wave signatures of the absence of an event horizon. II. Extreme mass ratio inspirals in the spacetime of a thin-shell gravastar},'' \href{http://dx.doi.org/10.1103/PhysRevD.81.084011}{{\em Phys. Rev. D} {\bfseries 81} (2010) 084011},
\href{http://arxiv.org/abs/1001.3031}{{\ttfamily arXiv:1001.3031 [gr-qc]}}.

\bibitem{Biswas:2023ofz}
S.~Biswas, C.~Singha, and S.~Chakraborty, ``{Galactic wormholes: Geometry, stability, and echoes},'' \href{http://dx.doi.org/10.1103/PhysRevD.109.064043}{{\em Phys. Rev. D} {\bfseries 109} no.~6, (2024) 064043}, \href{http://arxiv.org/abs/2307.04836}{{\ttfamily arXiv:2307.04836 [gr-qc]}}.

\bibitem{Abedi:2016hgu}
J.~Abedi, H.~Dykaar, and N.~Afshordi, ``{Echoes from the Abyss: Tentative evidence for Planck-scale structure at black hole horizons},'' \href{http://dx.doi.org/10.1103/PhysRevD.96.082004}{{\em Phys. Rev. D} {\bfseries 96} no.~8, (2017) 082004}, \href{http://arxiv.org/abs/1612.00266}{{\ttfamily arXiv:1612.00266 [gr-qc]}}.

\bibitem{Dey:2020lhq}
R.~Dey, S.~Chakraborty, and N.~Afshordi, ``{Echoes from braneworld black holes},'' \href{http://dx.doi.org/10.1103/PhysRevD.101.104014}{{\em Phys. Rev. D} {\bfseries 101} no.~10, (2020) 104014}, \href{http://arxiv.org/abs/2001.01301}{{\ttfamily arXiv:2001.01301 [gr-qc]}}.

\bibitem{Mark:2017dnq}
Z.~Mark, A.~Zimmerman, S.~M. Du, and Y.~Chen, ``{A recipe for echoes from exotic compact objects},'' \href{http://dx.doi.org/10.1103/PhysRevD.96.084002}{{\em Phys. Rev. D} {\bfseries 96} no.~8, (2017) 084002}, \href{http://arxiv.org/abs/1706.06155}{{\ttfamily arXiv:1706.06155 [gr-qc]}}.

\bibitem{Chakravarti:2021clm}
K.~Chakravarti, R.~Ghosh, and S.~Sarkar, ``{Signature of nonuniform area quantization on black hole echoes},'' \href{http://dx.doi.org/10.1103/PhysRevD.105.044046}{{\em Phys. Rev. D} {\bfseries 105} no.~4, (2022) 044046}, \href{http://arxiv.org/abs/2112.10109}{{\ttfamily arXiv:2112.10109 [gr-qc]}}.

\bibitem{Biswas:2022wah}
S.~Biswas, M.~Rahman, and S.~Chakraborty, ``{Echoes from braneworld wormholes},'' \href{http://dx.doi.org/10.1103/PhysRevD.106.124003}{{\em Phys. Rev. D} {\bfseries 106} no.~12, (2022) 124003}, \href{http://arxiv.org/abs/2205.14743}{{\ttfamily arXiv:2205.14743 [gr-qc]}}.

\bibitem{Biswas:2026zif}
S.~Biswas and S.~Chakrabarti, ``{Perturbations in the parametrized wormhole spacetime and their related quasinormal modes},'' \href{http://arxiv.org/abs/2605.05352}{{\ttfamily arXiv:2605.05352 [gr-qc]}}.

\bibitem{Shakura:1972te}
N.~I. Shakura and R.~A. Sunyaev, ``{Black holes in binary systems. Observational appearance},'' {\em Astron. Astrophys.} {\bfseries 24} (1973) 337--355.

\bibitem{Page:1974he}
D.~N. Page and K.~S. Thorne, ``{Disk-Accretion onto a Black Hole. Time-Averaged Structure of Accretion Disk},'' \href{http://dx.doi.org/10.1086/152990}{{\em Astrophys. J.} {\bfseries 191} (1974) 499--506}.

\bibitem{Novikov:1973kta}
I.~D. Novikov and K.~S. Thorne, ``{Astrophysics and black holes},'' in {\em {Les Houches Summer School of Theoretical Physics}: {Black Holes}}, pp.~343--550.
\newblock 1973.

\bibitem{Chakrabarti:1996cc}
S.~K. Chakrabarti, ``{Accretion processes on a black hole},'' \href{http://dx.doi.org/10.1016/0370-1573(95)00057-7}{{\em Phys. Rept.} {\bfseries 266} (1996) 229--392}, \href{http://arxiv.org/abs/astro-ph/9605015}{{\ttfamily arXiv:astro-ph/9605015}}.

\bibitem{Sharma:2026gvu}
C.~Sharma, K.~Chakravarti, and S.~Sarkar, ``{Quasinormal Ringdown and Echoes in Accreting Exotic Compact Objects},'' \href{http://arxiv.org/abs/2607.18919}{{\ttfamily arXiv:2607.18919 [gr-qc]}}.

\bibitem{Kotlarik:2018nbd}
P.~Kotla{\v{r}}{\'\i}k, O.~Semer{\'a}k, and P.~{\v{C}}{\'\i}{\v{z}}ek, ``{Schwarzschild black hole encircled by a rotating thin disc: Properties of perturbative solution},'' \href{http://dx.doi.org/10.1103/PhysRevD.97.084006}{{\em Phys. Rev. D} {\bfseries 97} no.~8, (2018) 084006}, \href{http://arxiv.org/abs/1804.02010}{{\ttfamily arXiv:1804.02010 [gr-qc]}}.

\bibitem{Kotlarik:2022spo}
P.~Kotla\v{r}\'\i{}k and D.~Kofro\v{n}, ``{Black Hole Encircled by a Thin Disk: Fully Relativistic Solution*},'' \href{http://dx.doi.org/10.3847/1538-4357/ac9620}{{\em Astrophys. J.} {\bfseries 941} no.~1, (2022) 25}, \href{http://arxiv.org/abs/2211.04823}{{\ttfamily arXiv:2211.04823 [gr-qc]}}.

\bibitem{Chen:2023akf}
C.-Y. Chen and P.~Kotla\v{r}\'\i{}k, ``{Quasinormal modes of black holes encircled by a gravitating thin disk},'' \href{http://dx.doi.org/10.1103/PhysRevD.108.064052}{{\em Phys. Rev. D} {\bfseries 108} no.~6, (2023) 064052}, \href{http://arxiv.org/abs/2307.07360}{{\ttfamily arXiv:2307.07360 [gr-qc]}}.

\bibitem{Kuzmin}
G.~Kuzmin, ``A stationary galaxy model admitting triaxial velocity distribution,'' {\em Astron. zh} {\bfseries 33} (1956) 27.

\bibitem{1963ApJ...138..385T}
A.~{Toomre}, ``{On the Distribution of Matter Within Highly Flattened Galaxies.},'' \href{http://dx.doi.org/10.1086/147653}{{\em \apj} {\bfseries 138} (Aug., 1963) 385}.

\bibitem{1993MNRAS.265..126B}
J.~{Bicak}, D.~{Lynden-Bell}, and C.~{Pichon}, ``{Relativistic Discs and Flat Galaxy Models},'' \href{http://dx.doi.org/10.1093/mnras/265.1.126}{{\em \mnras} {\bfseries 265} (Nov., 1993) 126}.

\bibitem{1978A&A....63..221A}
M.~{Abramowicz}, M.~{Jaroszynski}, and M.~{Sikora}, ``{Relativistic, accreting disks.},'' {\em \aap} {\bfseries 63} (Feb., 1978) 221--224.

\bibitem{Olivares-Sanchez:2024dfh}
H.~R. Olivares-S{\'a}nchez, P.~Kocherlakota, and C.~A.~R. Herdeiro, ``{GRMHD Simulations of~Accretion Onto Exotic Compact Objects},'' \href{http://arxiv.org/abs/2408.09893}{{\ttfamily arXiv:2408.09893 [astro-ph.HE]}}.

\bibitem{Stephani:2003tm}
H.~Stephani, D.~Kramer, M.~A.~H. MacCallum, C.~Hoenselaers, and E.~Herlt, \href{http://dx.doi.org/10.1017/CBO9780511535185}{{\em {Exact solutions of Einstein's field equations}}}.
\newblock Cambridge Monographs on Mathematical Physics. Cambridge Univ. Press, Cambridge, 2003.

\bibitem{1992MNRAS.257..152E}
N.~W. {Evans} and P.~T. {de Zeeuw}, ``{Potential-density pairs for flat galaxies},'' \href{http://dx.doi.org/10.1093/mnras/257.1.152}{{\em \mnras} {\bfseries 257} no.~1, (July, 1992) 152--176}.

\bibitem{2009MNRAS.396.1487V}
D.~{Vogt} and P.~S. {Letelier}, ``{Analytical potential-density pairs for flat rings and toroidal structures},'' \href{http://dx.doi.org/10.1111/j.1365-2966.2009.14803.x}{{\em \mnras} {\bfseries 396} no.~3, (July, 2009) 1487--1498}, \href{http://arxiv.org/abs/0906.0919}{{\ttfamily arXiv:0906.0919 [astro-ph.GA]}}.

\bibitem{Chen:2022ynz}
C.-Y. Chen, H.-W. Chiang, and J.-S. Tsao, ``{Eikonal quasinormal modes and photon orbits of deformed Schwarzschild black holes},'' \href{http://dx.doi.org/10.1103/PhysRevD.106.044068}{{\em Phys. Rev. D} {\bfseries 106} no.~4, (2022) 044068}, \href{http://arxiv.org/abs/2205.02433}{{\ttfamily arXiv:2205.02433 [gr-qc]}}.

\bibitem{goldberger2004effective-1ce}
W.~D. Goldberger and I.~Z. Rothstein, ``{An Effective field theory of gravity for extended objects},'' \href{http://dx.doi.org/10.1103/PhysRevD.73.104029}{{\em Phys. Rev. D} {\bfseries 73} (2006) 104029}, \href{http://arxiv.org/abs/hep-th/0409156}{{\ttfamily arXiv:hep-th/0409156}}.

\bibitem{thorne1980multipole-8ba}
K.~S. Thorne, ``Multipole expansions of gravitational radiation,'' \href{http://dx.doi.org/10.1103/revmodphys.52.299}{{\em Reviews of Modern Physics} {\bfseries 52} no.~2, (1980) 299--339}.

\bibitem{Radulea:2026hub}
A.~Radulea, M.~E. Rubio, K.~Kritos, and A.~Maselli, ``{Orbital evolution of asymmetric binaries within accreting environments},'' \href{http://dx.doi.org/10.1103/mk8r-8db5}{{\em Phys. Rev. D} {\bfseries 114} no.~6, (2026) 063059}, \href{http://arxiv.org/abs/2606.18341}{{\ttfamily arXiv:2606.18341 [gr-qc]}}.

\bibitem{Bhowmik:2026owi}
A.~Bhowmik, A.~Chowdhury, and S.~Chakrabarti, ``{Ringdown and tide: Fingerprints of dark matter halo profiles},'' \href{http://dx.doi.org/10.1103/mvys-wjf4}{{\em Phys. Rev. D} {\bfseries 114} no.~8, (2026) 084070}, \href{http://arxiv.org/abs/2608.07678}{{\ttfamily arXiv:2608.07678 [gr-qc]}}.

\bibitem{Ghosh:2026fbr}
R.~Ghosh, N.~Speeney, A.~Chowdhury, C.~Singha, M.~A.~S. Pinto, and E.~Berti, ``{From Density to Mass: A New Parametrized Framework for Dark Matter Environment Around Black Holes},'' \href{http://arxiv.org/abs/2609.35771}{{\ttfamily arXiv:2609.35771 [gr-qc]}}.

\bibitem{Vogt:2009sy}
D.~Vogt and P.~S. Letelier, ``{Analytical Potential-Density Pairs for Flat Rings and Toroidal Structures},'' \href{http://dx.doi.org/10.1111/j.1365-2966.2009.14803.x}{{\em Mon. Not. Roy. Astron. Soc.} {\bfseries 396} (2009) 1487}, \href{http://arxiv.org/abs/0906.0919}{{\ttfamily arXiv:0906.0919 [astro-ph.GA]}}.

\end{thebibliography}\endgroup
\bibliographystyle{./utphys1}
\end{document}